\documentclass[reprint,amsmath,amssymb,aps]{revtex4-1}

	\usepackage{graphicx}
	\usepackage{xcolor}
	\usepackage{dcolumn}
	\usepackage{bm}
	\usepackage{hyperref}

\newcommand{\nn}{\nonumber}

\begin{document}

\title{Quantum implications of non-extensive statistics}
\author{Nana Cabo Bizet}
 \email{nana@fisica.ugto.mx}
\affiliation{Departamento de F\'isica, Divisi\'on de Ciencias e Ingenier\'ias, Universidad de Guanajuato\\
Loma del Bosque 103, Le\'on 37150, Guanajuato, M\'exico.}
\author{C\'esar Dami\'an Ascencio }
\email{cesar.damian@ugto.mx}
\affiliation{Departamento de Ingenier\'{\i}a Mec\'anica, Carretera Salamanca - Valle de Santiago km 3.5 + 1.8, Comunidad de Palo Blanco,  Salamanca, Guanajuato, M\'exico. }
\author{Octavio Obreg\'on}
\email{octavio@fisica.ugto.mx}
 \affiliation{Departamento de F\'isica, Divisi\'on de Ciencias e Ingenier\'ias, Universidad de Guanajuato\\
Loma del Bosque 103, Le\'on 37150, Guanajuato, M\'exico.}
\author{Roberto Santos-Silva} 
\email{roberto.santos@academicos.udg.mx}
\affiliation{Departamento de Ciencias Naturales y Exactas, CUValles, Universidad de Guadalajara. Carretera Guadalajara-Ameca Km. 45.5, C.P. 46600, Ameca, Jalisco, M\'exico.}

%


\begin{abstract}
Exploring the analogy between quantum mechanics and statistical mechanics
we formulate an integrated version of the Quantropy functional \cite{baez}.
With this prescription we compute the propagator associated to Boltzmann-Gibbs statistics 
in the semiclassical approximation as $K=F(T) \exp\left(i S_{cl}/\hbar\right)$.
We determine also propagators associated to different non-additive statistics;
those are the entropies depending only on the probability  $S_{\pm}$ \cite{Obregon:2010zz} and Tsallis entropy $S_q$ \cite{tsallis}. 
For $S_{\pm}$ we obtain a power series solution for the probability vs. the energy, which
can be analytically continued to the complex plane, and employed to obtain the propagators. 
Our work is motivated by \cite{TsallisQM1} where a modified q-Schr\"odinger
equation is obtained; that provides the wave function for the free particle as a q-exponential. 
The modified q-propagator obtained with our method, leads to the same q-wave function for that case. 
The procedure presented in this work allows to calculate q-wave functions in problems
with interactions; determining non-linear quantum implications of non-additive statistics. In a similar manner the corresponding generalized wave functions associated to $S_{\pm}$ can also be constructed. The corrections to the original propagator are explicitly determined in the case of a free particle and the harmonic
oscillator for which the semi-classical approximation is exact.

\begin{description}
\item[PACS numbers]  03.65.-w; 05.70.Ln; 05.90.+m.
\item[keywords]
quantropy, non linear quantum systems, propagator, non-extensive entropies. \end{description}

\end{abstract}

\maketitle

\newpage
\section{Introduction}

Non-extensive entropies depending only on the probabilities have been obtained in \cite{Obregon:2010zz}. They belong to
a family of non-extensive statistical mechanics, relevant for non-equilibrium systems. Renowned examples are: R\'enyi \cite{renyi},
Tsallis ($S_q$) \cite{tsallis,TsallisQM1} and Sharma-Mital \cite{sharma}, all of them can be
obtained in the frame of Superstatistics \cite{super}. 

For the entropies depending only on the probability there are two entropy functionals :
\begin{eqnarray}
S_+=\sum_{l}(1-p_1^{p_l}), \,  \,  \,  \,   \, S_-=\sum_{l}(p_l^{-p_l}-1),\nonumber
\end{eqnarray}
which can be considered as building blocks for non-extensive entropies without parameters. For example one
can consider $S_1=(S_++S_-)/2$. These entropies are noticeably distinct to Boltzmann-Gibbs(BG) entropy for systems with few degrees of freedom; however when the number of degrees of freedom 
goes to the thermodynamical limit, they match perfectly BG statistics \cite{jesus19}. They belong
to the class of Superstatistics \cite{super} with an intensive parameter $\chi^2$  distribution  \cite{Obregon:2010zz}.

There is a universality of the Superstatistics family \cite{super}.   As it has been shown: for several distributions of the temperature 
the Boltzmann factor essentially coincides up to the first expansion terms. This has as a consequence
that also the entropies associated to these Boltzmann factors have all of them basically 
the same first corrections to the usual entropy. Furthermore, the three entropies listed
here that depend only on the probability are expanded only on the parameter
$y=p\ln p$, this is always smaller than 1 giving correction terms 
to the entropy which at any order are smaller than the previous ones.
So, that any function of $y$ proposed as another
generalized entropy, depending only on this parameter,
when expanded in $y$ will basically coincide 
with one of the three ones studied here.
Clearly demanding that the first term in the expansion
is $-y$ giving BG entropy. 
Thus the entropies $S_+$ , $S_-$, and their linear combination 
can be considered as building blocks to compute any possible modified entropies depending only
on the probability. 

Motivated by the concept of Quantropy
developed by Baez and Pollard \cite{baez} and by non linear quantum systems with modified wave functions based on Tsallis  statistics
\cite{TsallisQM1,TsallisQM2,QMT}, we develop a version of
Quantropy in terms of  the propagator of a quantum mechanical theory. The work \cite{Chavanis19}, appeared recently, it studies non linear quantum equations with 
the harmonic oscillator potential. Our generalized propagators could possibly be connected
to the appropriate quantum equations.  Baez and Pollard's Quantropy is a functional of the amplitude on the path integral $a$, with the same functional form as the entropy in terms of
the probability $Q=-\int_X a(x) \ln a(x)dx$. Giving the functional:
\begin{align}
\Phi_{BP}&=-\int_X a(x) \ln a(x)dx-\alpha \int_X a(x) dx \nn \\
&-\lambda\int_X a(x) S(x) dx.\label{BPQ}
\end{align}
From the search of extrema of this functional, restricted to values of $a$ normalized and an average of the action,
Baez and Pollard obtain the relation $a= \exp(i S/\hbar-1-\alpha)$ with $\lambda=\frac{1}{i \hbar}$. 
Then $a\sim \exp(i S/\hbar)$ with the normalization fixed by the Lagrange multiplier $\alpha$.

In their approach the energy is mapped to the action $S$ and the temperature to $i \hbar$.
We consider the same identification, but instead we identify $E$ with the classical
action $S_{cl}$. Thus we consider as the analogue of the entropy a functional in terms of the propagator, instead
of the amplitude $a$. This
is an extrapolation of the Quantropy \cite{baez} to an integration over all classical paths. The standard expression for the propagator
 is given semi-classically  by  $K\sim \exp(i S_{cl}/\hbar)$. We use this fact to define a kind of integrated Quantropy functional now
 in terms of the propagator for BG statistics, which we extend to the modified statistics $S_+$, $S_-$ and $S_q$.

This paper is organized as follows: In Section \ref{power} we obtain a series expansion for the probabilities
versus $\beta E$ for the generalized entropies depending only on the probabilities $S_+$ and $S_-$.
In Section \ref{series} we extend this expansions to the complex plane. In Section \ref{IQu} we present a version of
Quantropy for BG statistics, $S_+$ and $S_-$ and $S_q$.
In Section \ref{freeProp} we study in particular the case of the free particle propagator, obtained
from the extrema of the Quantropy in the cases of $S_+$ and $S_-$  and
$S_q$ for $q<1$ and $q>1$. We show that the $K_q$ propagator results exactly in the q-exponential that defines
the q-wave function for the free particle \cite{TsallisQM1}.  In a similar manner we argue
that the corresponding generalized propagators $K_+$ and $K_-$ provide us with a procedure
to construct $\Psi_+$ and $\Psi_-$ for the free particle. Our method however gives
the possibility to construct $K_q$, $K_+$ and $K_-$ also for problems with interactions
and by this mean to identify the corresponding wave functions. We also provide a way to
perform the normalization inspired in Feynman and Hibbs work \cite{FeynmanHibbs}. Section
\ref{oscProp} is devoted to the analysis of the harmonic oscillator, we exemplify
with the case corresponding to $S_+$. In Section \ref{conclusion} we summarize our results and
present the conclusions. An Appendix \ref{app} describes a numerical study
of the propagators.

\section{Probability distributions for systems with maximal $S_+$ and $S_-$}
\label{power}

We start by developing a recurrent solution for the probability distribution of the generalized entropy $S_+$ \cite{Obregon:2010zz}. 
On the contrary to BG statistics, for a system subject to $S_+$ extremization, probability normalization and energy conservation 
there is not a simple inverse function of the probabilities $p_l$  v.s. the energy of the state $\beta E_l$. We overcome this difficulty by
finding a series solution to the extremum equation.  There are other possible series solutions, but we discuss here one that has a good convergence. 
At the end of the section we give also the probability expansion for the entropy $S_-$, which is obtained by an
equivalent Ansatz.

The functional to maximize in terms of $S_+$ entropy and for a certain distribution of probabilities $p_l$ is given by \cite{PhysRevE.88.062146,obregon15}
\begin{eqnarray}
\Phi_+=\sum_l \left(1-p_l^{p_l}\right)-\gamma \sum_l p_l-\beta\sum_l E_lp_l^{p_l+1},
\end{eqnarray}
$\beta$ and $\gamma$ are Lagrange multipliers, and $E_l$ is the energy of the state $l$. The average values of energy and the normalization value have been omitted
for simplicity. 
This gives a relation between the energy and the probabilities, which we now denote $p$ without the index:
\begin{eqnarray}
\beta E=\frac{(-\gamma p^{-p} - 1 - \ln p)}{(1 + p + p \ln p)}.\label{ecu1}
\end{eqnarray}
Setting the Lagrange multiplier $\gamma$ to $-1$, we first expand the previous equation around $p=0$. 
That accounts to consider the expansion around $y=p \ln p=0$.
\begin{align}
e^{-\beta E}&=p - p^2 \ln p^2 + 1/2 p^3 (\ln p^2 + 2 \ln p^3 + \ln p^4) +\nn \\
&1/6 p^4 (-3 \ln p^2 - 8 \ln p^3 - 9 \ln p^4 - 6 \ln p^5 - \ln p^6)  \nn\\
&1/24 p^5 (12 \ln p^2 + 44 \ln p^3 + 70 \ln p^4 + 68 \ln p^5 \nn\\
& + 42 \ln p^6 +12 \ln p^7 + \ln p^8) +... \, \, . \label{exp1}
\end{align}

\begin{figure}[htbp]
\begin{center}
\includegraphics[width=.4\textwidth]{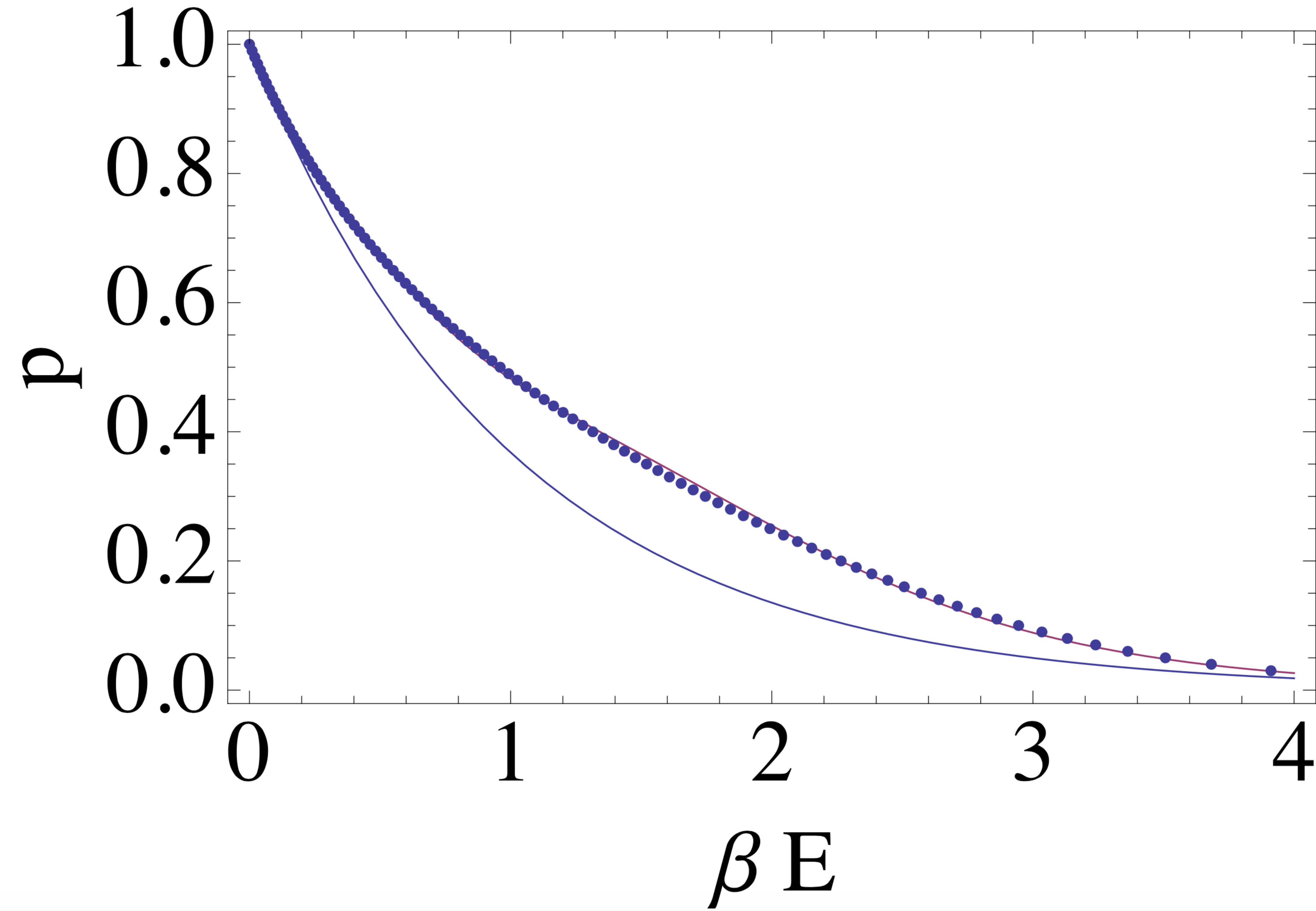}
\caption{Probability versus $\beta E$. The blue line represents the BG statistics distribution.
The dots represents the exact dependence in (\ref{ecu1}), $S_+$ statistics distribution,
while the red line represents the power series solution (\ref{sol1}) till  orden 3. i.e. up to the $e^{-4 x}$ 
correction.}
\label{figura1}
\end{center}
\end{figure}

We make the following Ansatz to solve the equation (\ref{exp1})
\begin{equation}
p=e^{-\beta E}(1+\sum_{n=1} c_n e^{-n \beta E}).\label{px}
\end{equation}
Plugging (\ref{px}) in (\ref{exp1}), to have the l.h.s. equal to the r.h.s. the coefficients multiplying the powers of $e^{-n \beta E}$ with $n>1$ have
to vanish.
This gives  a recurrent expression for the coefficients $c_n$, which for the first
four coefficients is solved as
\begin{align}
c_1&= x^2,\nn \\
 c_2&=1/2 x^2 (-1 - 2 x + 3 x^2),\nn \\
c_3&=\frac{1}{6} x^2 (3 + 4 x - 6 x^2 - 24 x^3 + 16 x^4), \nn \\
c_4&=\frac{1}{24} x^2 \left(-12 - 16 x + 60 x^2 + 116 x^3 \right.\nn \\
&\left.+ 30 x^4 - 300 x^5 + 125 x^6\right). \label{coef}
\end{align}
We have denoted $\beta E$ as $x$. The coefficients (\ref{coef}) give the following approximate solution for the probabilities versus $\beta E$
\begin{align}
p_+&=e^{-x} + e^{-2x} x^2 + \frac{1}{2}e^{-3 x} x^2 (-1 - 2 x + 3 x^2) \label{sol1} \\
&+\frac{1}{6}e^{-4 x} x^2 (3 + 4 x - 6 x^2 - 24 x^3 + 16 x^4) \nn \\
&+\frac{1}{24} e^{-5 x} x^2 \left(-12 - 16 x + 60 x^2 + 116 x^3 \right.\nn \\
&\left(+ 30 x^4 - 300 x^5 + 125 x^6\right))+... \,  \, .\nn
\end{align}
In Figure \ref{figura1} we compare the exact value of $p$ vs. $\beta E$ with the power series solution
(\ref{sol1}) till 3rd order, and with the Boltzmann distribution $e^{-\beta E}$.

\subsubsection*{Entropy $S_-$}

Consider the other generalized entropy dependent only on the probabilities is $S_-$. For this entropy, defining
the functional $\Phi_-$
as
\begin{eqnarray}
\Phi_-=\sum_l \left(p_l^{-p_l}-1\right)-\gamma \sum_l p_l-\beta\sum_l E_lp_l^{1-p_l}.\label{PhiMinus}
\end{eqnarray}
Finding the extrema of (\ref{PhiMinus})  and proposing the same Ansatz (\ref{px}) we obtain
a set of equations that can be solved to give the recursive probability solution:
\begin{align}
p_-&=e^{-x} (1 - e^{-x} x^2 + \frac{1}{2} e^{-2 x} x^2 (-1 - 2 x + 3 x^2) + \nn \\
   &\frac{1}{6} e^{-3 x} (-3 x^2 - 4 x^3 + 6 x^4 + 24 x^5 - 16 x^6))+...  \, \, \, .
\end{align}

\section{Modified amplitude expansions}
\label{series}

In this Section we use the series solutions for the probabilities in terms of the energy obtained in the previous Section, to perform an analytic continuation
into the complex plane. Consider $a$ as the amplitude of a path, new complex variable substituting the probability $p_l$, 
and $A$ as the action replacing $\beta E_l$. This will allow to study modified quantropy functionals,
for the definition of Baez and Pollard (\ref{BPQ}), as well as for our definition (\ref{Phi+}). The usual Quantropy
solution will give an exponential $a\sim e^{\frac{i A}{\hbar}}$. In our approach this would be $K\sim e^{\frac{i S_{cl}}{\hbar}}$. We want to analyze the new statistics
$S_+$ and $S_-$. We will find a functional dependence of $a$ vs. $A$ ($K$ vs. $S_{cl}$) 
to deviate from the exponential.

The main idea is to complexify the power expansion solution (\ref{sol1}) since the amplitude is a complex number, such
that we have a solution to the extrema of the modified Quantropy
\begin{align}
\Phi_{BP,+}&=\int \left(1-a(x)^{a(x)}\right) dx-\alpha\int a(x) dx \nn \\
&-\lambda\int A(x)a(x)^{a(x)+1} dx.\label{Phi+A}
\end{align}
Finding the extrema of (\ref{Phi+A}) w.r.t to $a$ one gets
\begin{eqnarray}
\frac{A}{i\hbar}=\frac{(-\gamma a^{-a} - 1 - \ln a)}{(1 + a + a \ln a)}=F\left(a\left(\frac{A}{i\hbar}\right)\right).\label{ecuI}
\end{eqnarray}
The range of validity of the propagators computations depends on the convergence of the imaginary series
solution to this equation. The series is obtained by doing the replacement $\beta E$ by $\frac{A}{i\hbar}$  and $p$ by $a$ in (\ref{ecu1}). Also the Lagrange multipliers have to me mapped as $\beta$ to $\lambda$ and $\alpha$ to $\gamma$. The minus sign gives the right sign after a rotation on the argument of the exponential (similar to a Wick rotation).  As the series solution continuation of (\ref{sol1}) we obtain the following expression 

\begin{align}
\label{amplitud}
a_+\left(\frac{A}{i\hbar}\right)&=e^{iA/\hbar} \left( 1- \frac{A}{\hbar}^2 e^{iA/\hbar} \right. \\
&\left.-\frac{1}{2} A^2/\hbar^2 (-1+2iA/\hbar -3A^2/\hbar^2) e^{i2A/\hbar}\right. \nn \\
&-\left.\frac{1}{6} A^2/\hbar^2 (3-4iA/\hbar + 6A^2/\hbar^2 \right.\nn \\
&\left. -24iA^3/\hbar^3 + 16A^4/\hbar^4)e^{3iA/\hbar} +...\right).\nonumber
\end{align}
Since $A$ has units of action the argument of the exponentials and the terms on the expansion are adimensional. Substituting this expression on the constraint equation (\ref{ecuI}) we obtain the real and imaginary parts of $F(a\left(\frac{A}{i\hbar}\right))$.
Beginning in Figure 3, the relevant difference of the propagators $K_+$ obtained
with (\ref{amplitud}) with respect to the standard one can be observed in the region of  $S_{cl}\approx \hbar$.

Using the parameter $\lambda = \frac{1}{i \hbar}$ in ($\ref{amplitud}$) the expression becomes 

\begin{align}
\label{amplitud1}
a_+(A)&= e^{-\lambda A} \left( 1+ (\lambda A)^2 e^{-\lambda A}\right. \nn\\
&\left.+ \frac{1}{2} (\lambda A)^2 (-1-2 \lambda A + 3(\lambda A)^2) e^{-2\lambda A} \right. \nonumber \\ 
&\left.  +\frac{1}{6} (\lambda A)^2 (3 + 4 \lambda A - 6 (\lambda A)^2\right.\nn \\
&\left. -24(\lambda A)^3 + 16 (\lambda A)^4)e^{-3 \lambda A} +...\right).
\end{align}
Is not difficult to observe that any term of this expansion can be written as derivatives with respect to the parameter $\lambda$. 
If we derive with respect to $\lambda$ the usual amplitude we obtain:
 $\frac{\partial}{\partial \lambda}e^{-\lambda A} = - A e^{- \lambda A}$. Higher derivates 
 can be written as
\begin{equation}
\left( \frac{\lambda}{n}\right)^m \frac{\partial^m}{\partial \lambda^m} e^{-n \lambda A} = (-1)^m (\lambda A)^m e^{-n\lambda A} ,
\end{equation}
where $m$ and $n$ are positive integers. Thus we rewrite ($\ref{amplitud1}$) as

\begin{align}
a_+&=e^{-\lambda A}  + \frac{\lambda^2}{4} \frac{\partial^2}{ \partial \lambda^2} e^{-2\lambda A} \label{aMas} \\
& + \frac{1}{2} \left( - \frac{\lambda^2}{3^2} \frac{\partial^2}{ \partial \lambda^2} + 2 \frac{\lambda^3}{3^3} \frac{\partial^3}{ \partial \lambda^3} + 3\frac{\lambda^4}{3^4} \frac{\partial^4}{\partial \lambda^4} \right)e^{-3 \lambda A}   \nn \\ 
& +\frac{1}{6} \left(3 \frac{\lambda^2}{4^2} \frac{\partial^2}{\partial \lambda^2} - 4 \frac{\lambda^3}{4^3} \frac{\partial^3}{\partial \lambda^3} - 6 \frac{\lambda^4}{4^4} \frac{\partial^4}{\partial \lambda^4} + 24\frac{\lambda^5}{4^5} \frac{\partial^5}{\partial \lambda^5}+\right.\nn\\
&\left. 16 \frac{\lambda^6}{4^6} \frac{\partial^6}{\partial \lambda^6}\right)e^{-4 \lambda A}+... \, \, \,  .\nn
\end{align}
One can compute the corrections to any order. In the case of $a_0=e^{i S/\hbar}$ those corrections can be interpreted as higher order interactions 
of the action at different frequencies of the usual amplitude \footnote{For example for a massive particle those will be contributions from multiples of the particle mass. For the harmonic oscillator also there will be contributions with a tower of masses and frequencies.}.

Now one can apply the same method to determine the distribution arising from the  Quantropy  with statistics $S_-$.
We also have to perform the extension to the complex plane. The distribution for the modified Quantropy coming 
from $S_-$  is given by:
\begin{align}
a_-&=e^{i A/\hbar} \left(1 + (A/\hbar)^2 e^{i A/\hbar}\right. \nn \\
 &\left.- \frac{1}{2} (A/\hbar)^2 (-1 + 2 i A/\hbar - 3 (A/\hbar)^2) e^{2 i A/\hbar} \right.  \nn \\
&\left.+ \frac{1}{6} (3 (A/\hbar)^2 - 4 i (A/\hbar)^3 + 6 (A/\hbar)^4 - 24 i (A/\hbar)^5\right. \nn \\
&\left.+ 16 (A/\hbar)^6) e^{3  i(A/\hbar)}+...\right). 	\label{aMenos}\end{align}

\section{Quantropy in terms of the propagator}

\label{IQu}
In this section we present as an alternative proposal a kind of integrated version of the Quantropy \cite{baez}. First we do it for the BG entropy, then for 
$S_+,S_-$ and $S_q$. The change in distribution probabilities which arise from modified entropies in statistics
is now reflected in the quantum arena as modifications to the propagators. The propagators
between points in space-time  $(\mathbf{x}_a,t_a)$ and $(\mathbf{x}_b,t_b)$ in quantum mechanics determine the probability 
amplitude of particles to travel from certain position
to another position in a given time. As modified entropies in statistical physics
lead to modified probability distributions, distinct probabilities of propagation over all paths 
from $(\mathbf{x}_a,t_a)$ and $(\mathbf{x}_b,t_b)$ will arise from a modified Quantropy.

In the work \cite{baez} the Quantropy functional associated with BG statistics was formulated,  and its maximization leads
to the weight on the path integral $a\sim \exp(-\lambda S)$ with $\lambda=\frac{1}{i\hbar}$.  We propose
another functional, which in a sense constitutes an integrated version of Quantropy.
Its maximization leads to the propagator $K(x)\sim \exp(-\lambda S_{cl}(x))$.
The  Wentzel-Kramers-Brillouin(WKB) method \cite{WKB} allows to compute the wave function in a semiclassical approximation. In a sense
this is linked to our approach, on which we maximize a functional which determines the propagator with a semiclassical approximation
in terms of the classical action $S_{cl}$. This is exact for the free particle, the harmonic oscillator
as well as for other cases \cite{SCA,SCA2}. The procedure is applied to generalized entropy functionals,
giving a modified propagator. For the Tsallis statistics we obtain $K_q(x)\sim\exp_q(-\lambda S_{cl}(x))$.
This structure is the same that the wave function for the free particle that the Tsallis
statistic possesses $\Psi_q(x)=\exp_q(i (k x-w t))$, which has been proposed
as solution to the non linear quantum equations of \cite{QMT}. According to Feynman arguments,
one can start with the free particle propagator and determine the corresponding wave function \cite{FeynmanHibbs}.
Thus our procedure allows to find a propagator which 
can be identified with the wave function of interest. We should note that the  propagator resulting
from our procedure will not only describe the free particle but to a good approximation any other problem
with its corresponding classical action.  Our method should give the wave function solution for the problem
of interest.

With the same method we write functionals for $S_+$, $S_-$ and obtain probability distributions,
we can write the corresponding Quantropies and obtain the propagators $K_+$ and $K_-$,
and correspondingly extrapolate them to the wave functions $\Psi_+$ and $\Psi_-$.
This would give us the quantum behavior for the corresponding action.
The given propagators, can be related to non linear quantum systems studied in the literature
\cite{Chavanis19}.

To define our functionals we use the semiclassical limit to compute the propagator, this is  $K(x)=F(a,b) e^{\frac{i S_{cl}(x)}{\hbar}}$, denoting the classical action  as $S_{cl}(x)$, and being $F(a,b)$ a constant depending on the time difference $t_b-t_a$. For the free particle and the harmonic oscillator as well as oder physical problems \cite{SCA,SCA2} this is  an exact result.

For the BG statistics we define the Quantropy functional
\begin{align}
\Phi_0&=-\int K(x)\ln K(x) dx-\alpha\int K(x) dx \nn\\
&-\lambda\int\left(S_{cl}(x)K(x) \right) dx.
\end{align}
The extrema condition $\frac{\delta \Phi_0}{\delta K(x)}=0$ gives as solution the propagator dependence $K(x)=e^{-1-\alpha-\lambda S_{cl}(x)}$, $\lambda=\frac{1}{i\hbar}$, where the normalization constant $\alpha$ determines $F(a,b)$.

The integrated Quantropy functional for the new $S_+$ statistic is given by
\begin{align}
\Phi_+&=\int \left(1-K(x)^{K(x)}\right) dx-\alpha\int K(x) dx \nn \\
&-\lambda\int\left(S_{cl}(x)\right) K(x)^{K(x)+1} dx.\label{Phi+}
\end{align}
The extrema condition $\frac{\delta \Phi_+}{\delta K(x)}=0$ gives the equation:
\begin{eqnarray}
\lambda S_{cl}(x)=\frac{-1-\ln K(x)-\alpha K(x)^{-K(x)}}{1+K(x)+\ln K(x)}.  \label{extrema}
\end{eqnarray}
Using our knowledge to resolve this type of equation from the statistical physics case, presented in Section \ref{power}, this gives for the modified propagator the series solution:
\begin{align}
K_+(x)&=N_+e^{-\lambda S_{cl}}\left(1-e^{-\lambda S_{cl}}(\lambda S_{cl})^2 \right.\label{kMas} \\
&\left. +e^{-2\lambda S_{cl}}(\lambda S_{cl})^2(-1-2(\lambda S_{cl})+3(\lambda S_{cl})^2) \right.
\nn\\
&\left.-\frac{1}{6} e^{-2 S_{cl} \lambda} \times(-3 S_{cl}^2 \lambda^2 - 
   4 S_{cl}^3 \lambda^3 + 6 S_{cl}^4 \lambda^4 +\right.\nn\\
   &\left. 24 S_{cl}^5 \lambda^5 - 
   16 S_{cl}^6 \lambda^6)+...\right). \nn
\end{align}
This is obtained by taking the normalization $\alpha=-1$. A different  normalization would change the coefficients
in the expansion (\ref{kMas}).

The maximization constraint for the new $S_-$ statistic is given by
\begin{align}
\Phi_-&=\int \left(K(x)^{-K(x)}-1\right) dx-\alpha\int K(x) dx\nn \\
&-\lambda\int\left(S_{cl}\right) K(x)^{-K(x)+1} dx.
\end{align}
The extrema condition $\frac{\delta \Phi_-}{\delta K(x)}=0$ gives the equation:
\begin{eqnarray}
\lambda S_{cl}(x)=\frac{1+\ln K(x)+\alpha K(x)^{K(x)}}{1-K(x)-\ln K(x)}.
\end{eqnarray}
Using our knowledge of this type of equation from the statistical physics case, we obtain for the modified propagator the series solution:
\begin{align}
K_-(x)&=N_-e^{-\lambda S_{cl}}\left(1+e^{(-\lambda S_{cl})}(\lambda S_{cl})^2 \right.\label{kMenos} \\
&\left.-e^{(-\lambda S_{cl})}(\lambda S_{cl})^2(-1-2(\lambda S_{cl})+3(\lambda S_{cl})^2\right.\nn \\
&\left.+\frac{1}{6} e^{-2 S_{cl} \lambda} (-3 S_{cl}^2 \lambda^2 - 
   4 S_{cl}^3 \lambda^3 + 6 S_{cl}^4 \lambda^4\right. \nn \\
  & \left. + 24 S_{cl}^5 \lambda^5 - 
   16 S_{cl}^6 \lambda^6)+...\right). \nn
\end{align}

In the case of Tsallis statistics the functional is given by:
\begin{eqnarray}
\Phi_q&=&\int \frac{\left(1-K(x)^q\right)}{(q-1)} dx-\alpha\int K(x) dx \nn \\
&-&\lambda\int\left(S_{cl}\right) K(x)^q dx,
\end{eqnarray}
and the solution is 
\begin{align}
K_q(x)&=N_q\exp_q(-\lambda S_{cl}(x)),\label{kT}\\
&=N_q\left(1-(1-q) \lambda S_{cl}\right)^{\frac{1}{1-q}}. \nn
\end{align}
We have still to discuss the normalization of the different Kernels. 
This q-propagator is related to the q-wave function for the free particle non
linear quantum mechanics of \cite{TsallisQM1}. We will specify to this case 
which has been studied by other means, in the literature \cite{TsallisQM1,QMT}.

\section{Free particle propagators}
\label{freeProp}

In this section we write a modified propagator up to third order for the free particle in the case of the statistics $S_+$, $S_-$ and $S_q$  for $q=1-\delta$ and $q=1+\delta$ with $\delta>0$. The values of $q$ less or equal than one are considered  in order to compare the different propagators. 
We determine when the corrections to the usual propagator play an important role, which turns to be in the quantum regime characterized by $S_{cl}\approx \hbar$.
First we describe the procedure, then the normalization and in the last subsection we summarize our results.

\subsection{Superposition of Kernels}

Now we proceed to describe a generalized Kernel. The generalized complex probability distribution given by the expansion  (\ref{kMas}), can be regarded as a superposition of Kernels. 
Furthermore, the superposition will carry to the wave functions. 
In order to normalize the superposition we consider that the total Kernel expansion integration is the same
of the usual (1 for the free particle),
as is explicit in the Quantropy functional (\ref{Phi+}).
We show that this coincides with the result for the normalization 
obtained from propagating the wave function \cite{FeynmanHibbs}.

For the free particle  the unnormalized Kernel is:
\begin{align}
K_0(x,t;0,0)=\left(\frac{2 \hbar \epsilon i \pi}{m}\right)^{(n-1)/2}\left(\frac{1}{n}\right)^{1/2}\exp\left(\frac{i m x^2}{2 \hbar t}\right). \nn
\end{align}
$n$ is the
number of divisions of the time interval and $\epsilon$ is an infinitesimal time parameter that satisfies $t=\epsilon n$. 
This expression arises from computing the path integral to get:
\begin{align} 
K_0(x,t;0,0)&= \int e^{i S/\hbar} D x \\
&= \int \exp\left(\frac{i m}{2 \hbar \epsilon}\sum_n (x_n-x_{n-1})^2\right)d^nx\nn \\
&= \left(\frac{i\pi}{2 A}\right)^{\frac{1}{2}}\left(\frac{2 i \pi}{3 A}\right)^{\frac{1}{2}}\left(\frac{3i \pi}{4 A}\right)^{\frac{1}{2}}\times \nn\\
&...\times\left(\frac{(n-1)i\pi}{n A}\right)^{\frac{1}{2}}\exp\left(\frac{i A (x_0-x_n)^2}{n}\right).\nn
\end{align}
with $A=\frac{m}{2\epsilon\hbar}$.  The normalization constant is given by $N=(\frac{2 \pi i \hbar \epsilon}{m})^{-\frac{n}{2}}$, hence the  normalized propagator  is 
\begin{eqnarray}
K_1(x,t;0,0)=\left(\frac{2 \hbar t i \pi}{m}\right)^{-1/2}\exp\left(\frac{i m x^2}{2 \hbar t}\right).
\end{eqnarray}

We define the unnormalized Kernel for the free particle as
\begin{eqnarray}
k(x,t;1)=\exp\left(\frac{i m x^2}{2 \hbar t}\right)=e^{-\lambda A},\label{unNorm}
\end{eqnarray}
and the first two corrections in $K_+$ are given by
\begin{align}
k(x,t;2)&=\left(\frac{m x^2}{\hbar t}\right)^2\exp\left(\frac{i m x^2}{\hbar t}\right)=(\lambda A)^2\frac{\partial^2}{\partial \lambda^2}e^{-\lambda A}, \nn\\
k(x,t;3)&=-\left(\frac{m^2 x^4}{8 \hbar^2 t^2} +\frac{m^3 x^6}{8 i \hbar^3 t^3} + 
  3 \frac{m^4 x^8}{32 \hbar^4 t^4} \right) \nn \\
  &\times\exp\left(\frac{3 i m x^2}{2 \hbar t}\right),\nn\\
  &= \frac{1}{2}\left( - \frac{\lambda^2}{3^2} \frac{\partial^2}{\partial \lambda^2} + 2 \frac{\lambda^3}{3^3} \frac{\partial^3}{\partial \lambda} + 3\frac{\lambda^4}{3^4} \frac{\partial^4}{\partial \lambda^4}\right) e^{-3 \lambda A}. \nn
  \end{align}
Thus the generalized Kernel associated with $S_+$ entropy is given by
\begin{eqnarray}
K_+(x,t)=N_+(k(x,t;1)+k(x,t;2)+k(x,t;3)+...).\nn
\end{eqnarray}
The normalization constant  is determined by the requirement $\int_{-\infty}^{\infty}K_+(x,t)dx=1$,
and up to the first corrections is given by  $N_{+}=\frac{1}{1 +3/(16 \sqrt{2})}=0.883...$. The
reason for this normalization is also understood by an argument presented in the following,
motivated by Feynmann and Hibbs procedure \cite{FeynmanHibbs} and discussed next.

Let us also discuss the normalization used in the modified propagators, as shown in the standard case \cite{FeynmanHibbs}. We start considering the original unnormalized Kernel for the free particle, computed from the path integral:
\begin{eqnarray}
K_{1,0}(x,t;0,0)= 2^{\frac{N-1}{2}} \left(\frac{\pi i \hbar t}{m N}\right)^{\frac{N-1}{2}} \left(N\right)^{-1/2}\exp \left( \frac{i m x^2}{2\hbar t}\right).\nn
\end{eqnarray}
To determine the normalization constant in the  Feynman and Hibbs method we can apply formulae (2-34) and (4-3) on their book \cite{FeynmanHibbs},
to write the new infinitesimal Kernel between position $x_i$ and $x_{i+1}$, with $\Delta x_i=x_{i+1}-x_{i}$, in a time $\epsilon$ as follows
\begin{align}
K_+(i_{i+1},i)&=\frac{1}{A} \exp\left(\frac{i \epsilon}{\hbar}L\left(\frac{\Delta x_i}{\epsilon},\frac{x_{i+1}+x_i}{2},\frac{t_{i+1}+t_i}{2}\right)\right),\nn\\
&\left(1+(i \epsilon/\hbar)^2 L\left(\frac{\Delta x_i}{\epsilon},\frac{x_{i+1}+x_i}{2},\frac{t_{i+1}+t_i}{2}\right)^2 \right. \nn \\
& \left. \exp(i \epsilon L(v,\bar x,\bar t)/\hbar))+...\right).\label{iKernel}
\end{align}
The method consists in writing the wave function at a position $x$ at a time $t+\epsilon$ in terms of the wave function at position $y=x+\eta$ at a time $t$, explicitly 
\begin{eqnarray}
\psi(x,t+\epsilon)&=&\int_{-\infty}^{\infty} K_+(x,y,\epsilon)\psi(y,t) dy, \label{psiGen}\\
&=&\int_{-\infty}^{\infty} \frac{1}{A} \exp\left(\frac{i \epsilon}{\hbar}L\left(\frac{x-y}{\epsilon},\frac{x+y}{2},t\right)\right) \nn \\
&&\times(1+...)\psi(y,t) dy,\nn\\
&=&\int_{-\infty}^{\infty} \frac{1}{A} \exp\left(\frac{i \epsilon}{\hbar}L\left(-\frac{\eta}{\epsilon},x+\frac{\eta}{2},t\right)\right)\nn \\
&&\times(1+...)\psi(x+\eta,t) d \eta,\nn \\
&=&\int_{-\infty}^{\infty} \frac{1}{A} \exp\left(\frac{i m\eta^2}{2\hbar\epsilon}\right) \exp\left(-\frac{i\epsilon V(x+\frac{\eta}{2},t)}{\hbar}\right)\nn \\
&&\times(1+...)\psi(x+\eta,t) d \eta.\nn
\end{eqnarray}
In the quantum standard theory the normalization constant can be determined by expanding the l.h.s, of (\ref{psiGen}) $\psi(x,t+\epsilon)=\psi(x,t)+\epsilon \partial_t \psi$, the r.h.s. $\psi(x+\eta)=\psi(x,t)+\eta \partial_x \psi+\frac{\eta^2}{2}\partial_x^2 \psi$ and $\exp (-i \epsilon V/\hbar)=1-\frac{i \epsilon V}{\hbar}+...$,  then we compare the leading $\epsilon^0$ term. This implies that
\begin{eqnarray}
\frac{1}{A_0} \int_{-\infty}^{\infty}\exp\left(\frac{i m\eta^2}{2\hbar\epsilon}\right)d\eta=1.
\end{eqnarray}
In a similarly fashion one would get for the first correction to $K_+$ written in (\ref{iKernel}) 

\begin{eqnarray}
\frac{1}{A} \int_{-\infty}^{\infty}\exp\left(\frac{i m\eta^2}{2\hbar\epsilon}\right) \left(1-\left(\frac{m\eta^2}{2\hbar \epsilon}\right)^2 e^{\frac{i m\eta^2}{2\hbar \epsilon}}+...\right)d \eta=1. \nn
\end{eqnarray}
 It is worth to mention that the more important contribution to (\ref{psiGen}) is given for small $\eta$'s, as well as in our generalized case.
 It is necessary to check this argument, in order to verify let us consider the following integrals:
 \begin{align}
\int e^{ i C w^2} dw&=\sqrt{i \pi/C},\, \, \int e^{ i C w^2} w^4 dw=\frac{3\sqrt{\pi}}{4(-i C)^{5/2}}, \nn \\
&\, \, \int e^{ i C w^2} w^{2n+1} dw=0,\, \, n\in \mathbb{N}.\nn
\end{align}
The first correction gives the relation:
\begin{eqnarray}
A=A_0\left(1+\frac{3}{16 \sqrt{2}}+...\right). \label{normA}
\end{eqnarray}
The previous normalization factor is a general feature to apply to any potential $V(x,t)$, in particular is valid for both cases discussed here:  the free particle and the harmonic oscillator. A similar expression holds for $K_-$ normalization, this will be calculated in next Section.

\subsection{Analysis of the propagators}

Here we summarize the propagators obtained with the normalization methods described in
previous subsections. The results for $K_+$ can be extrapolated to $K_-$ and
$K_q$ because the method applied to obtain all of the propagators is
basically the same.

Recall the standard propagator of the free particle from the space-time point $(0,0)$ to $(x,t)$ is given by
\begin{eqnarray}
K_1(x,t;0,0)& &=N_0 \exp\left(\frac{i m x^2}{2 \hbar t}\right).
\end{eqnarray}
The constant w.r.t. to x: $N_0=\sqrt{\frac{m}{2 \pi i \hbar t}}$. For 
the case of the $S_+$ and $S_-$ statistics the first two contributions to the modified propagators (\ref{kMas}) and (\ref{kMenos}) read: 
\begin{eqnarray}
K_{\pm}&=&N_{\pm}\exp\left(\frac{i m x^2}{2 \hbar t}\right)\times\\
&&\left(1\mp \exp\left(\frac{i m x^2}{2 \hbar t}\right)\left(\frac{i m x^2}{2 \hbar t}\right)^2+...\right). \nn
\end{eqnarray}
with $N_+\sim N_0$. For the Tsallis statics the associated propagator (\ref{kT}) is given by the expression:
\begin{eqnarray}
K_q(x)&=&N_q\left(1+(q-1)\left(\frac{m x^2}{2 \hbar i t}\right) \right)^{\frac{1}{1-q}}\\
&=&N_q\exp\left(\frac{i m x^2}{2 \hbar t}\right) \left(1-(q-1)\left(\frac{m^2 x^4}{2 \hbar^2 t^2}\right)+....\right).\nn
\end{eqnarray}
We calculate the normalization constants for $K_{\pm}$ up to the first correction and exactly
$K_q$ to get:
\begin{eqnarray}
N_+&=&\sqrt{\frac{m}{2 \pi i \hbar t}}\frac{1}{(1+\frac{3}{16 \sqrt{2}}+...)} ,\\
N_-&=& \sqrt{\frac{m}{2 \pi i \hbar t}}\frac{1}{(1-\frac{3}{16 \sqrt{2}}+...)} , \\
N_q&=&\sqrt{\frac{m}{2 \pi i \hbar t}} \frac{\sqrt{(q-1)}\Gamma(\frac{1}{(q-1)})}{\Gamma(\frac{1}{(q-1)}-\frac{1}{2})}.
\end{eqnarray}

\begin{figure}[htbp]
\begin{center}
\includegraphics[width=.2\textwidth]{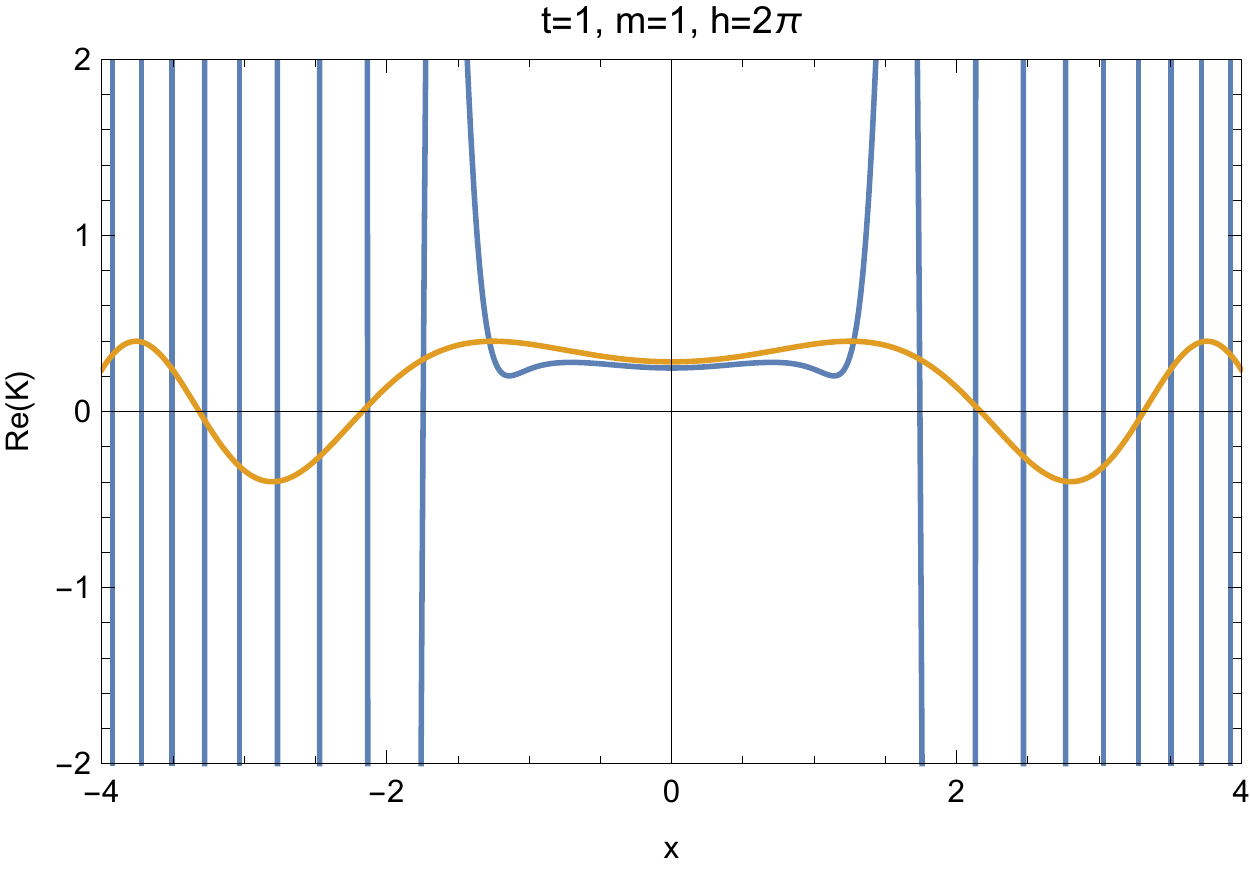}
\includegraphics[width=.2\textwidth]{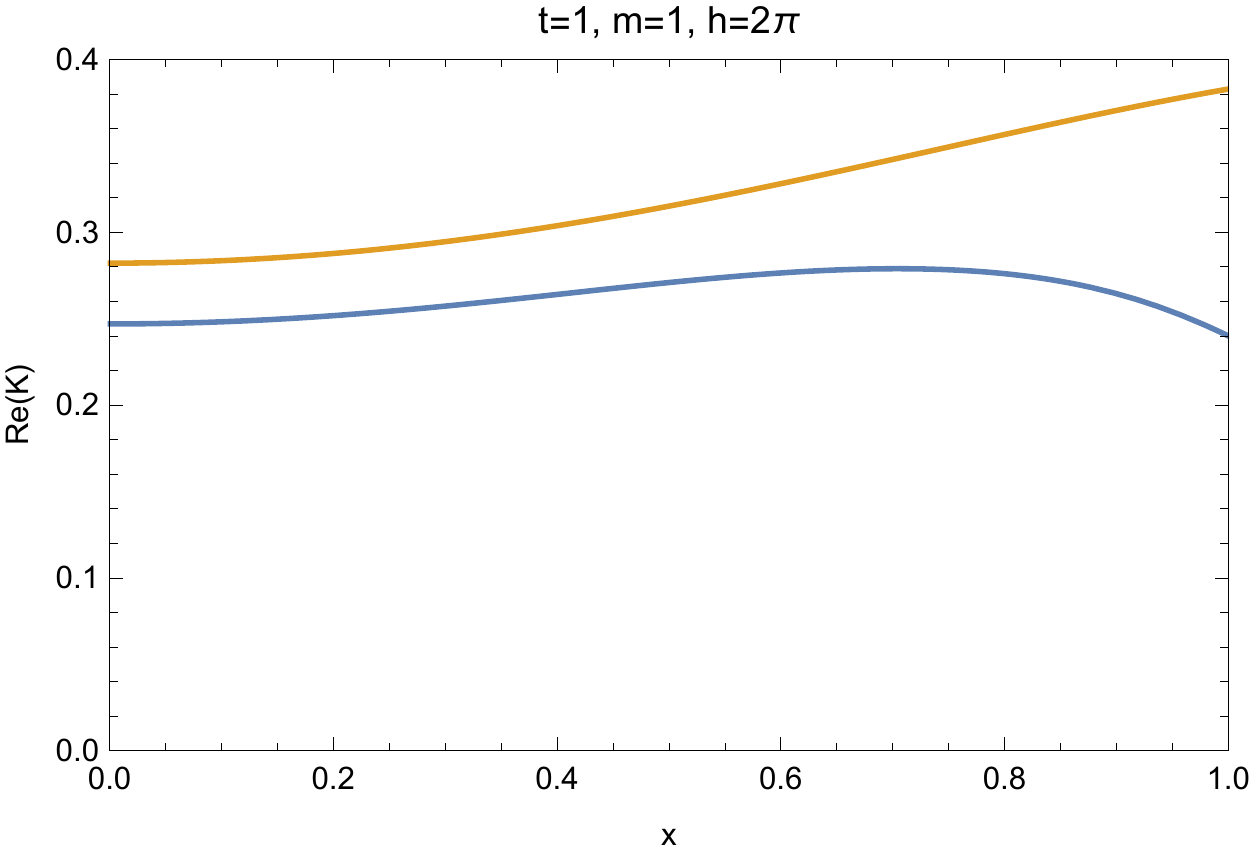}
\\
\includegraphics[width=.2\textwidth]{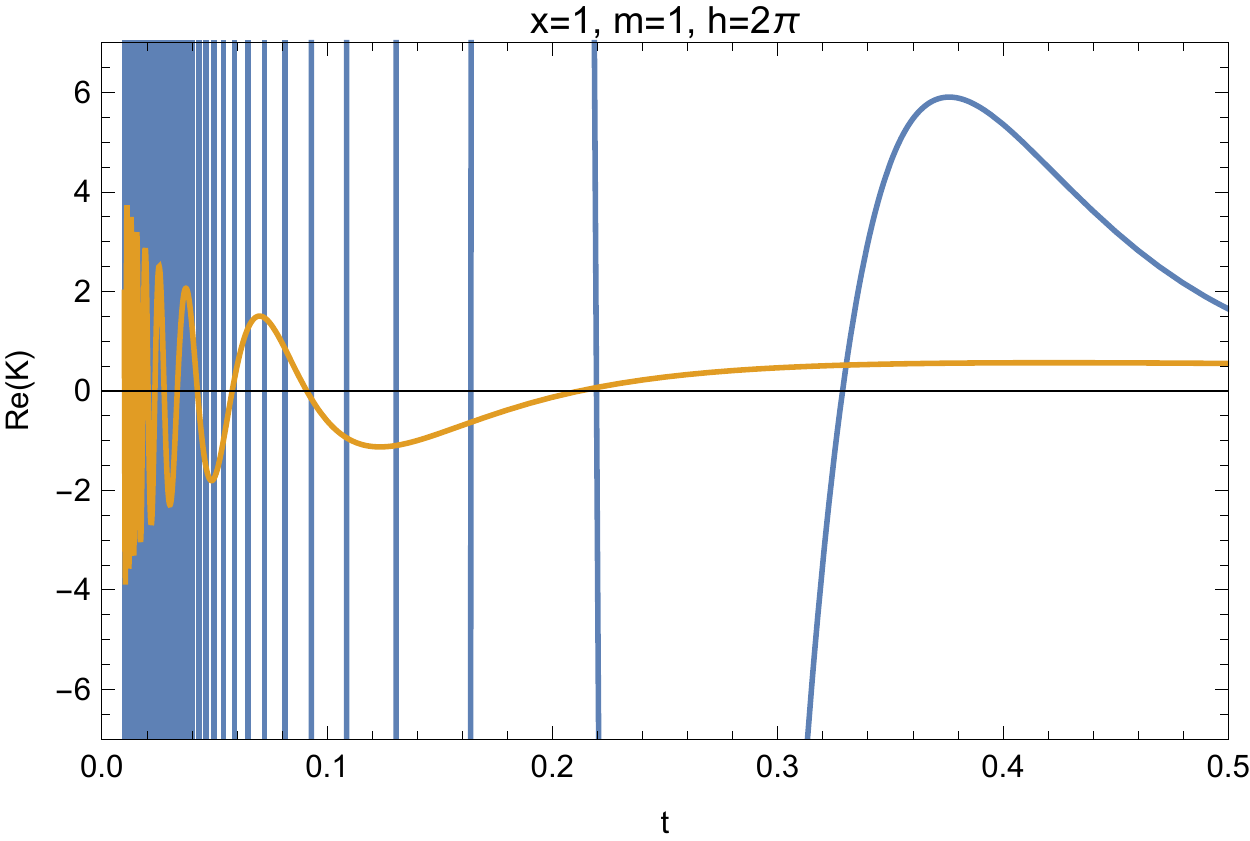}
\includegraphics[width=.2\textwidth]{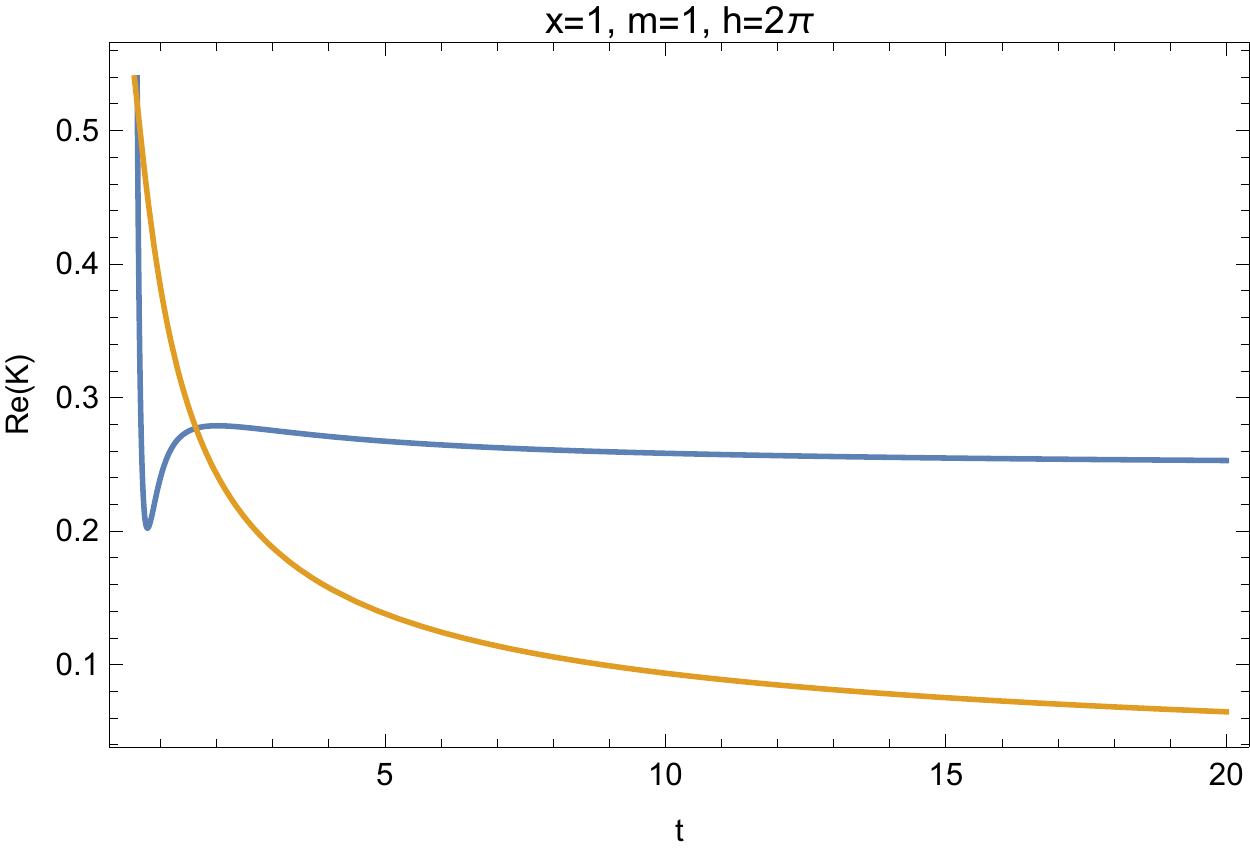}
\caption{Real parts of the modified propagator (blue line) vs. standard propagator (yellow line), for the free particle for the modified
statistics $S_+$. We set the mass and the Planck constant to unity. The quantum regime is given by $S_{cl} \approx \hbar$.
Imposing  $S_{cl} \lesssim \hbar$ which translates for fixed $x=1$ in $t \gtrsim1/2$, for fixed $t=1$ translates in $x^2 \lesssim 2$.  }
\label{quantum1}
\end{center}
\end{figure}

\begin{figure}[htbp]
\begin{center}
\includegraphics[width=.2\textwidth]{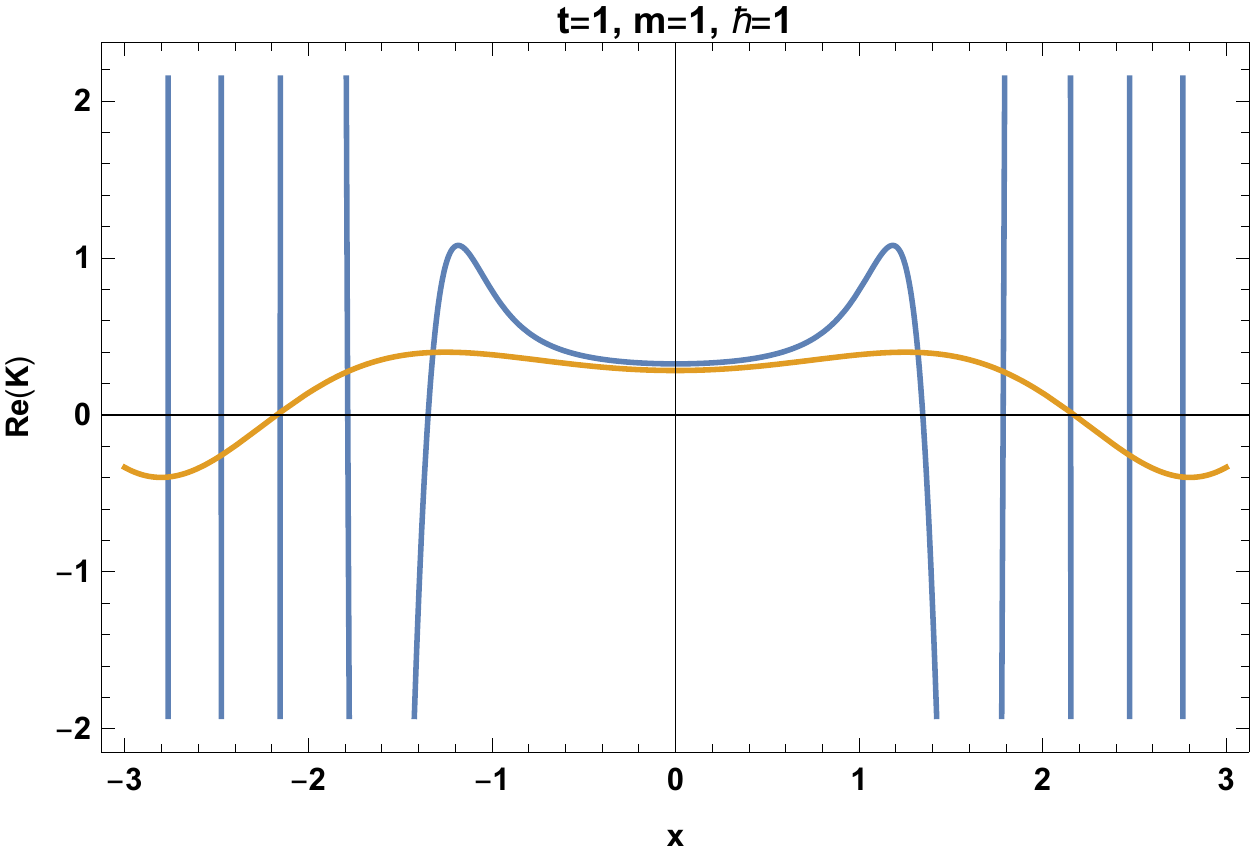}
\includegraphics[width=.2\textwidth]{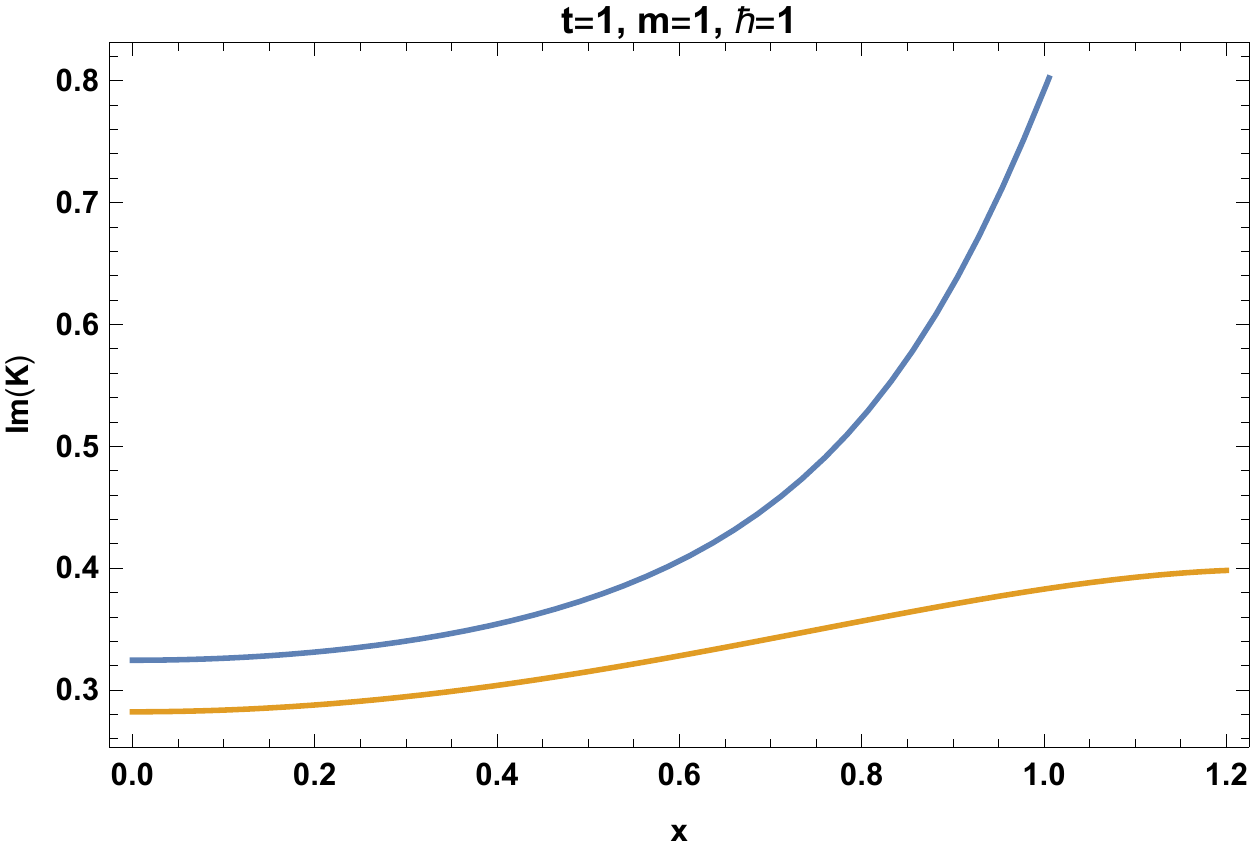}
\\
\includegraphics[width=.2\textwidth]{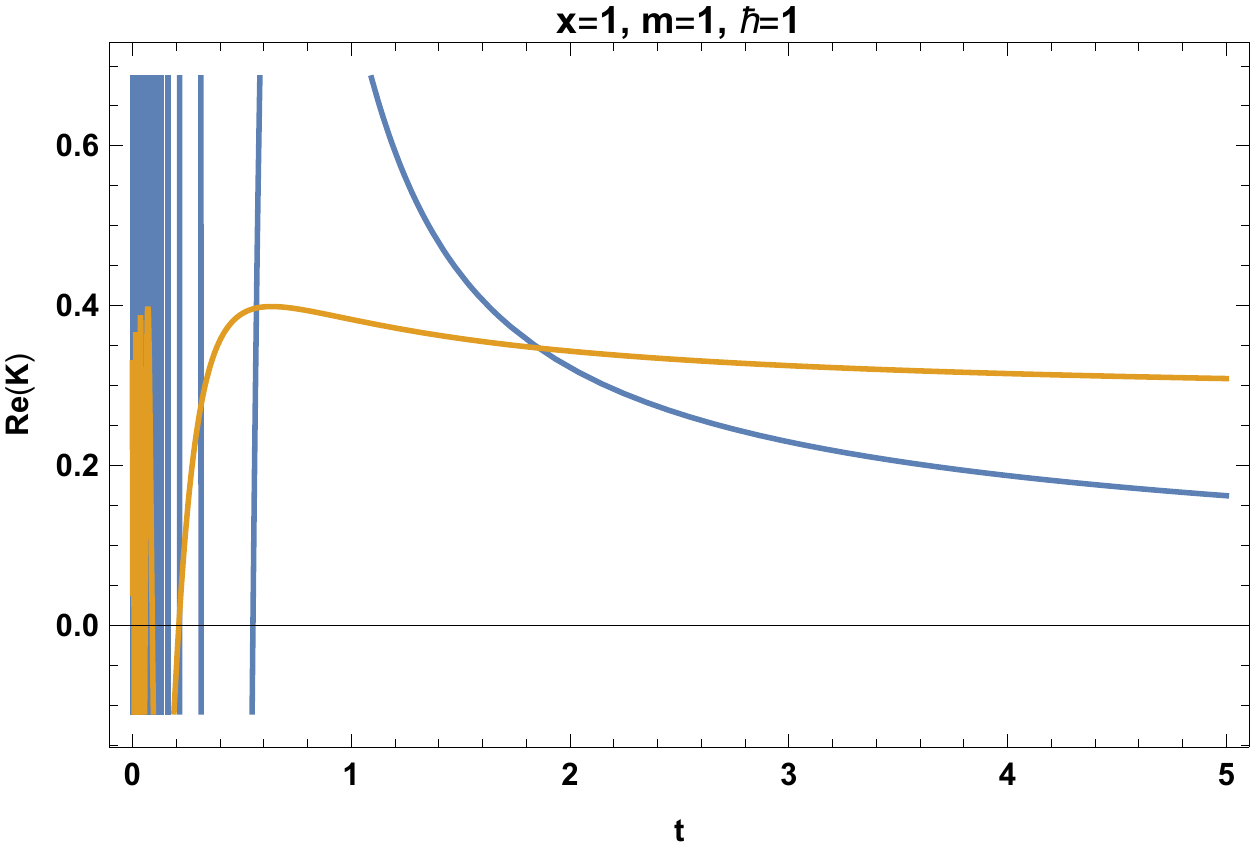}
\includegraphics[width=.2\textwidth]{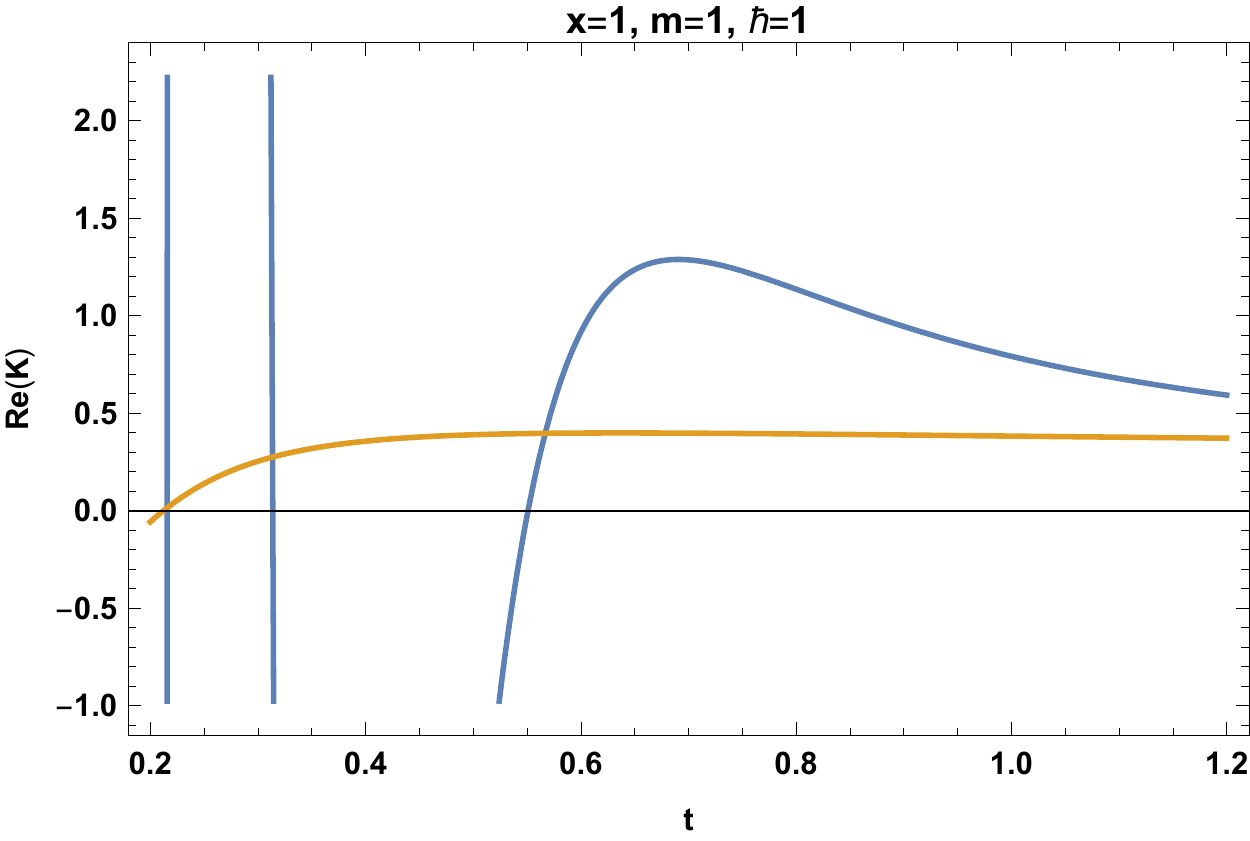}
\caption{Real parts of the modified propagator (blue line) vs. standard propagator (yellow line), for the free particle for the modified
statistics $S_-$. We set the mass and the Planck constant to unity. Imposing $S_{cl} \lesssim \hbar$ for fixed $x=1$ translates in $t \gtrsim1/2$, and for fixed $t=1$ translates in $x^2 \lesssim 2$.  }
\label{quantum2}
\end{center}
\end{figure}

\begin{figure}[htbp]
\begin{center}
\includegraphics[width=.2\textwidth]{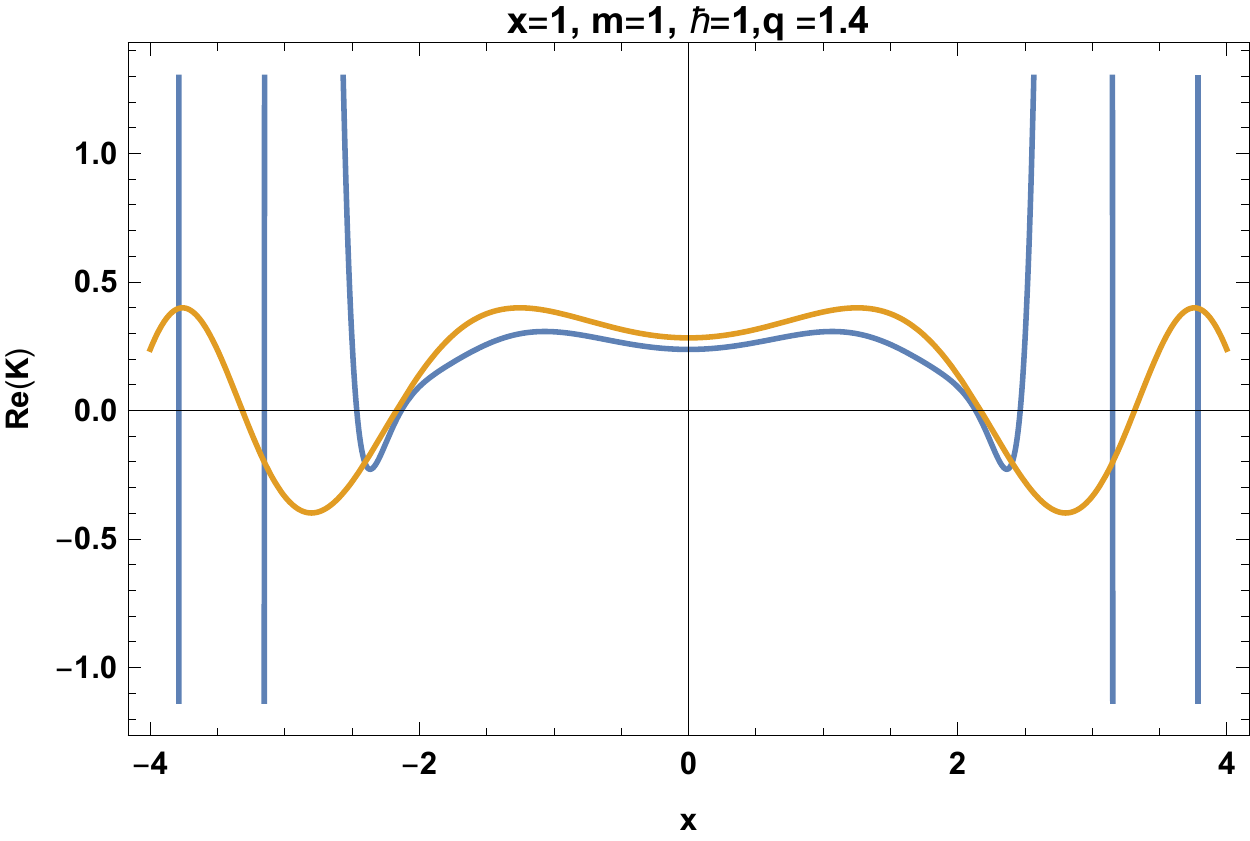}
\includegraphics[width=.2\textwidth]{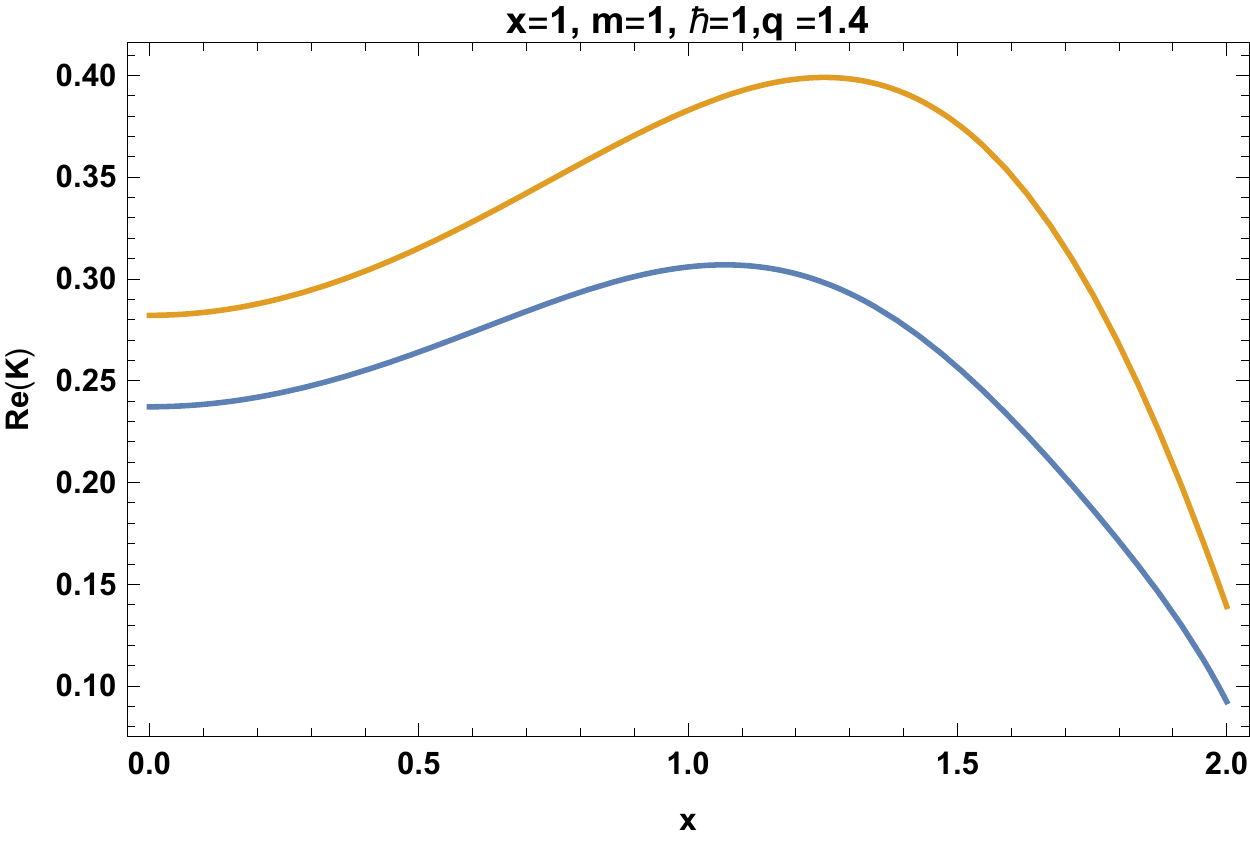}
\includegraphics[width=.2\textwidth]{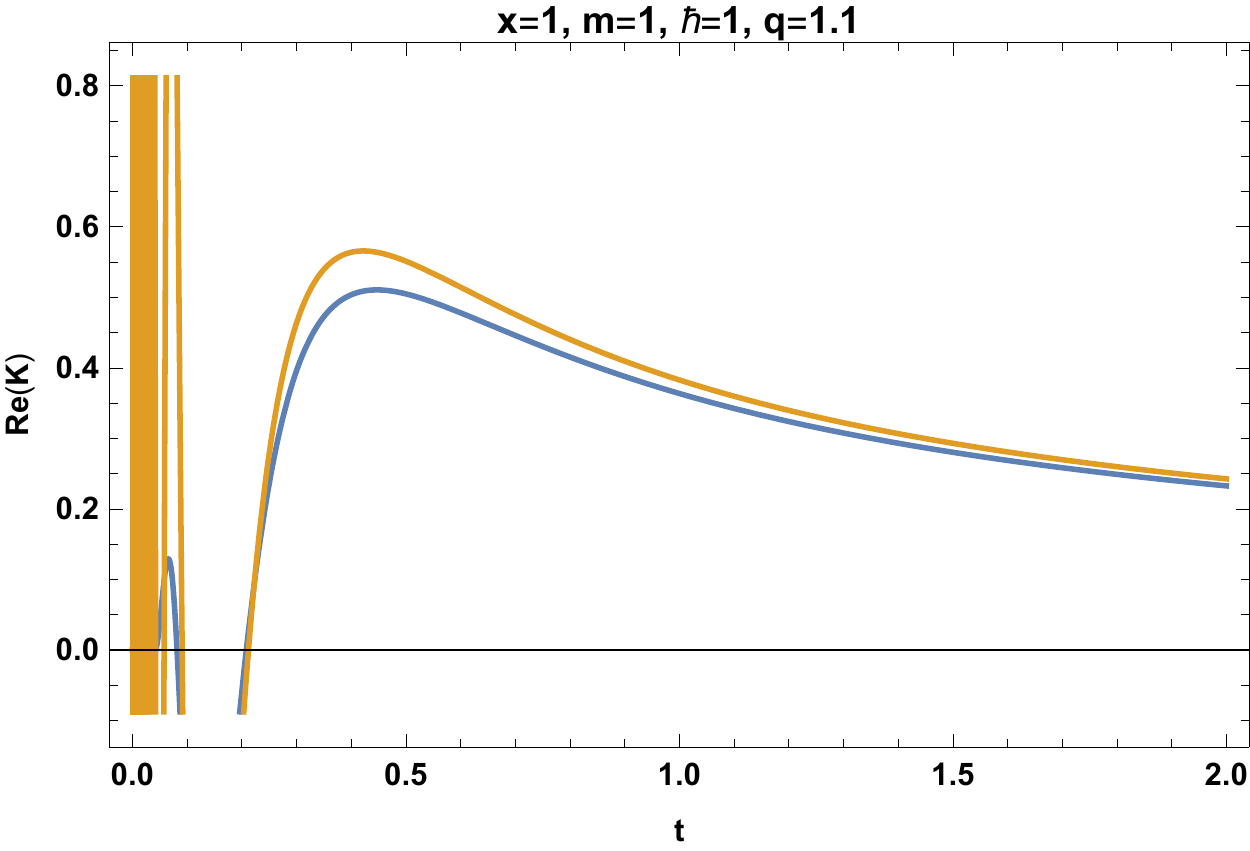}
\includegraphics[width=.2\textwidth]{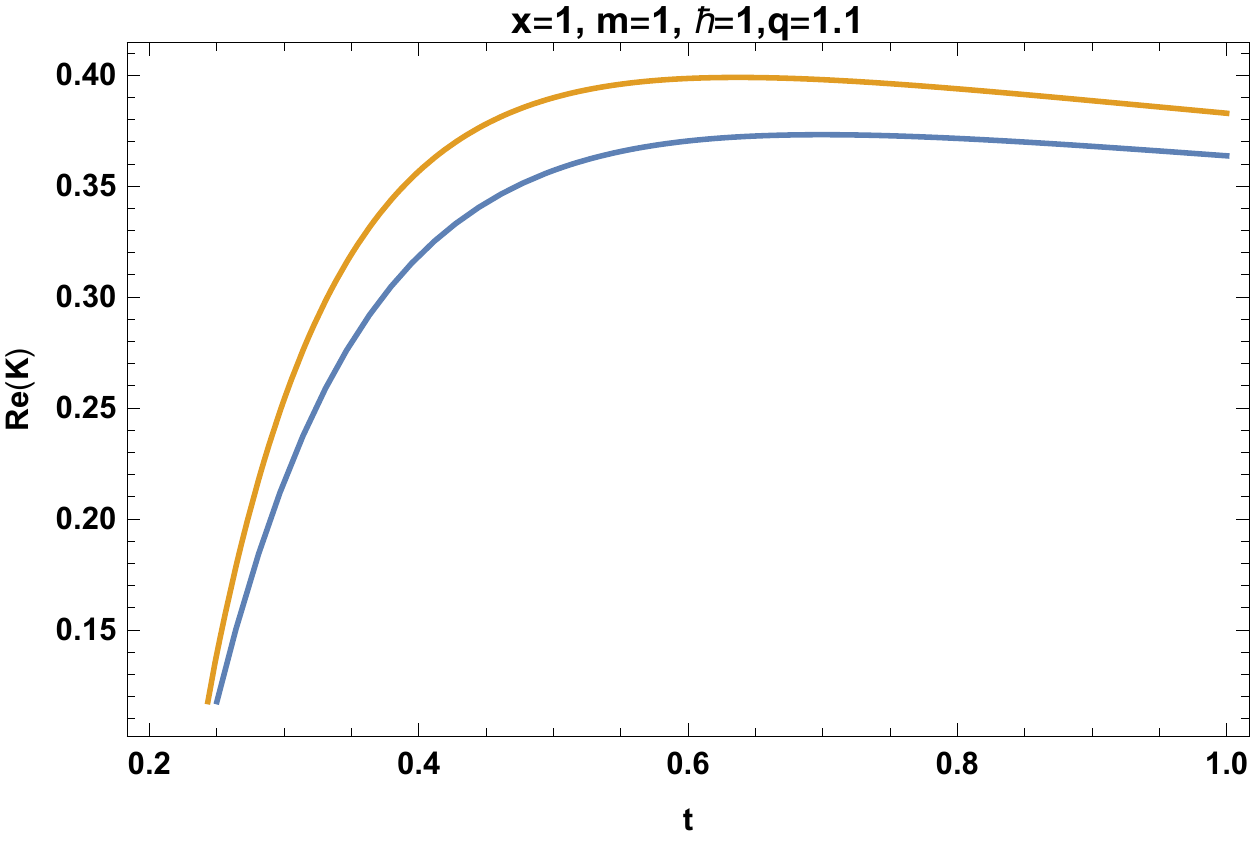}
\caption{Real parts of the modified propagator (blue line) vs. standard propagator (yellow line), for the free particle for the modified
statistics of Tsallis for $q=1.1$. We set the mass and the Planck constant to unity. Imposing $S_{cl} \lesssim \hbar$, and it translates for fixed $x=1$ in $t \gtrsim1/2$, for fixed $t=1$ translates in $x^2 \lesssim 2$.}
\label{quantumT2}
\end{center}
\end{figure}

In the quantum regime $S_{cl}\approx \hbar$ the differences between the propagators $K_+,K_-,K_q$ and $K_0$ are shown in Figures 
\ref{quantum1}, \ref{quantum2} and \ref{quantumT2}. Figure \ref{quantum1} compares $K_+$
 propagator with the usual one. Also Fig. \ref{quantum2} compares $K_-$
 propagator with the standard one. The last Figure \ref{quantumT2} does a comparison between
 the propagators for the $S_q$ statistics  $K_q$ for $q<1$
 and $q>1$ with the usual one. 
 The region of interest is the quantum regime with $S_{cl}\approx \hbar$.
  Furthermore in the classical regime the oscillations of the standard propagator
 grow averaging to zero \cite{FeynmanHibbs}. Thus we are interested in comparing the corrections arising from
different statistics in the quantum region of interest.

\begin{figure}[htbp]
\begin{center}
\includegraphics[width=.2\textwidth]{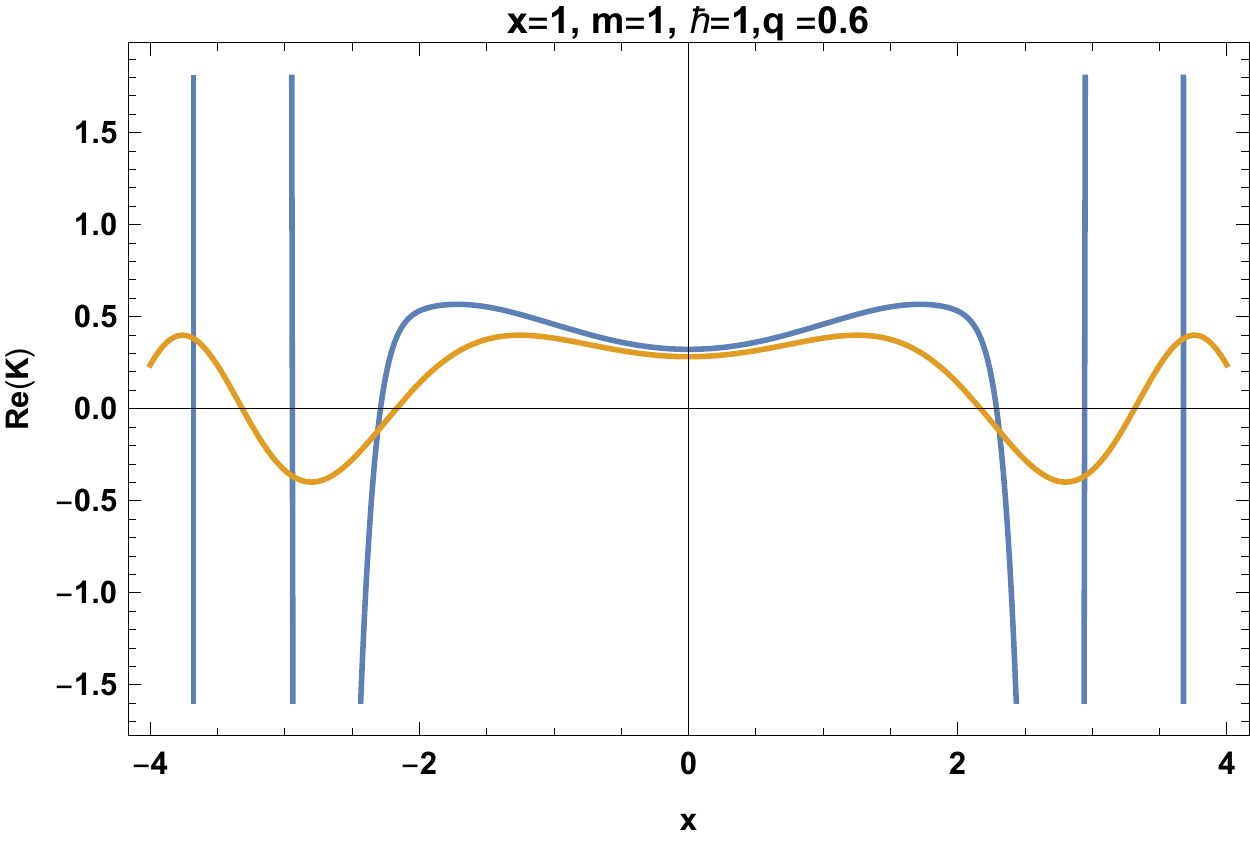}
\includegraphics[width=.2\textwidth]{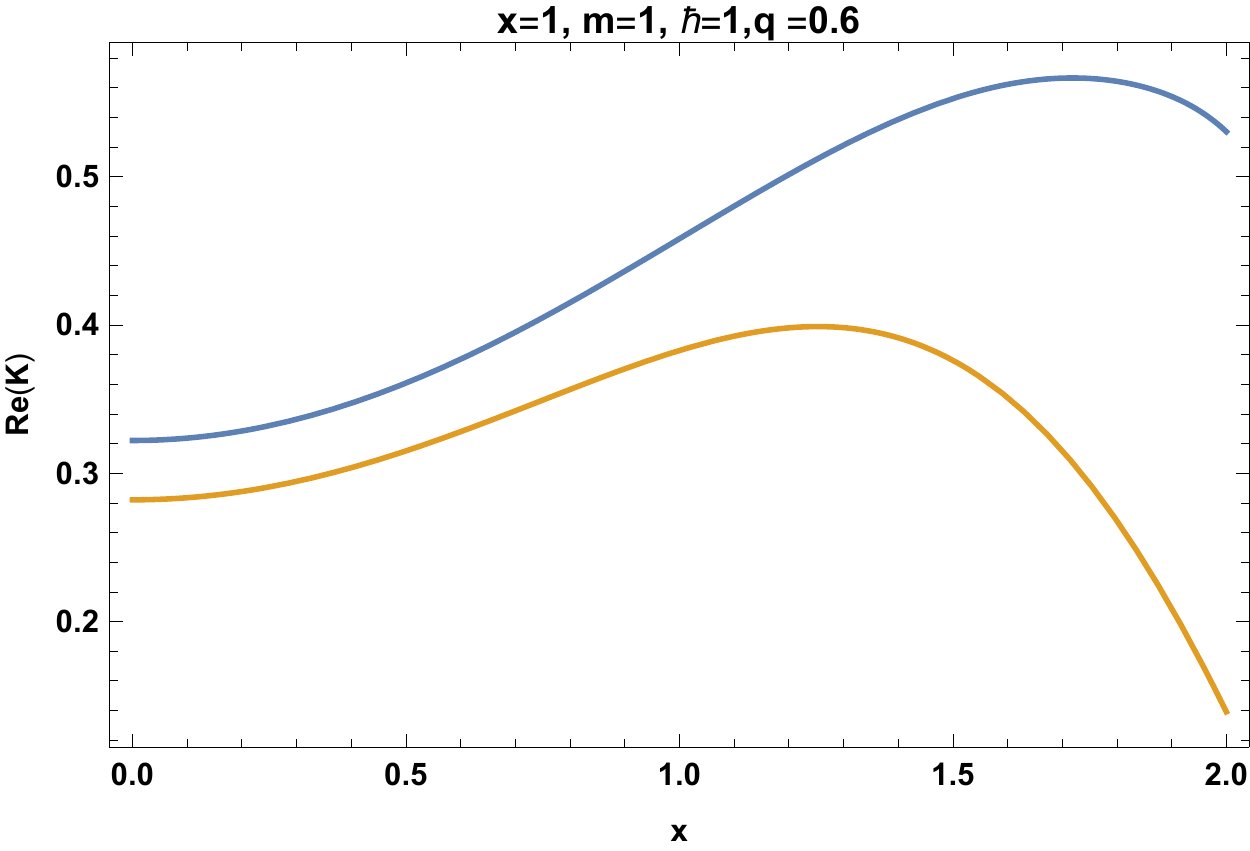}
\includegraphics[width=.2\textwidth]{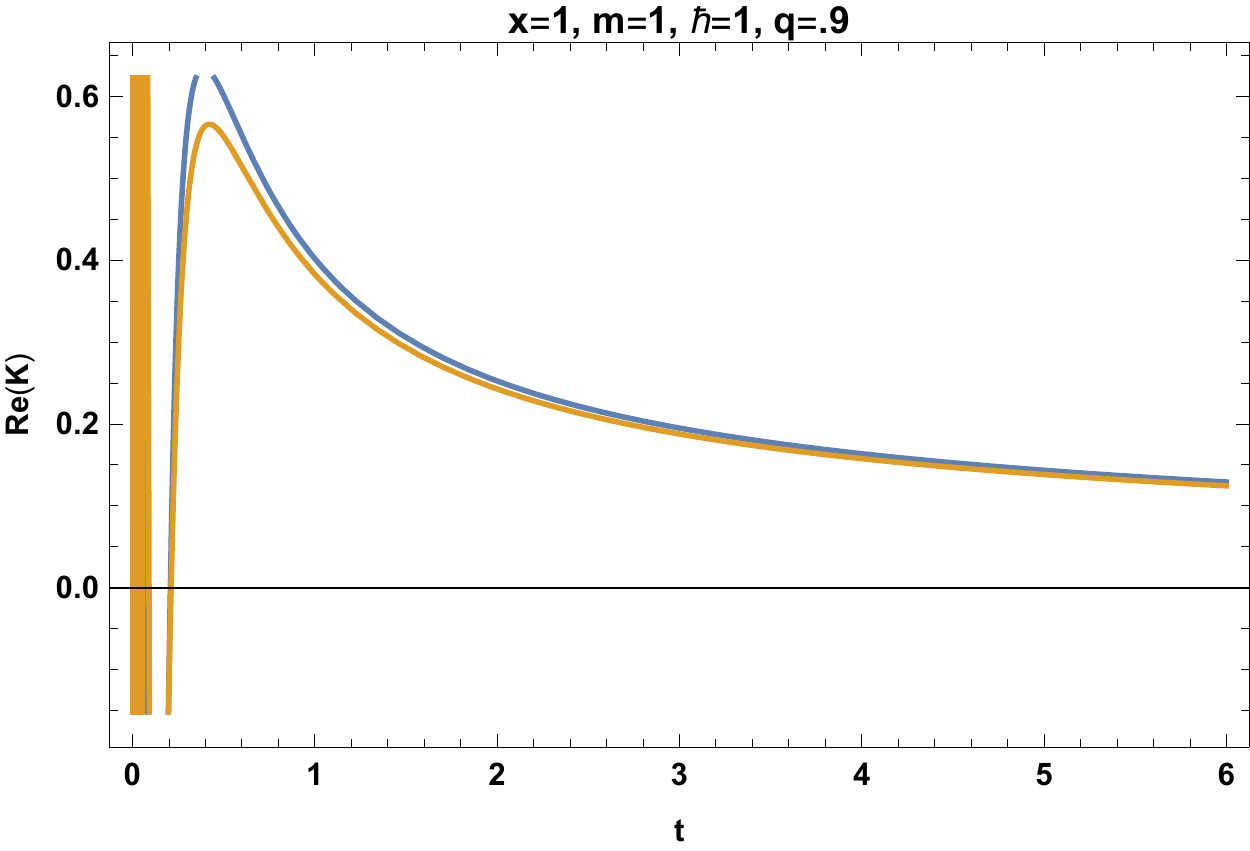}
\includegraphics[width=.2\textwidth]{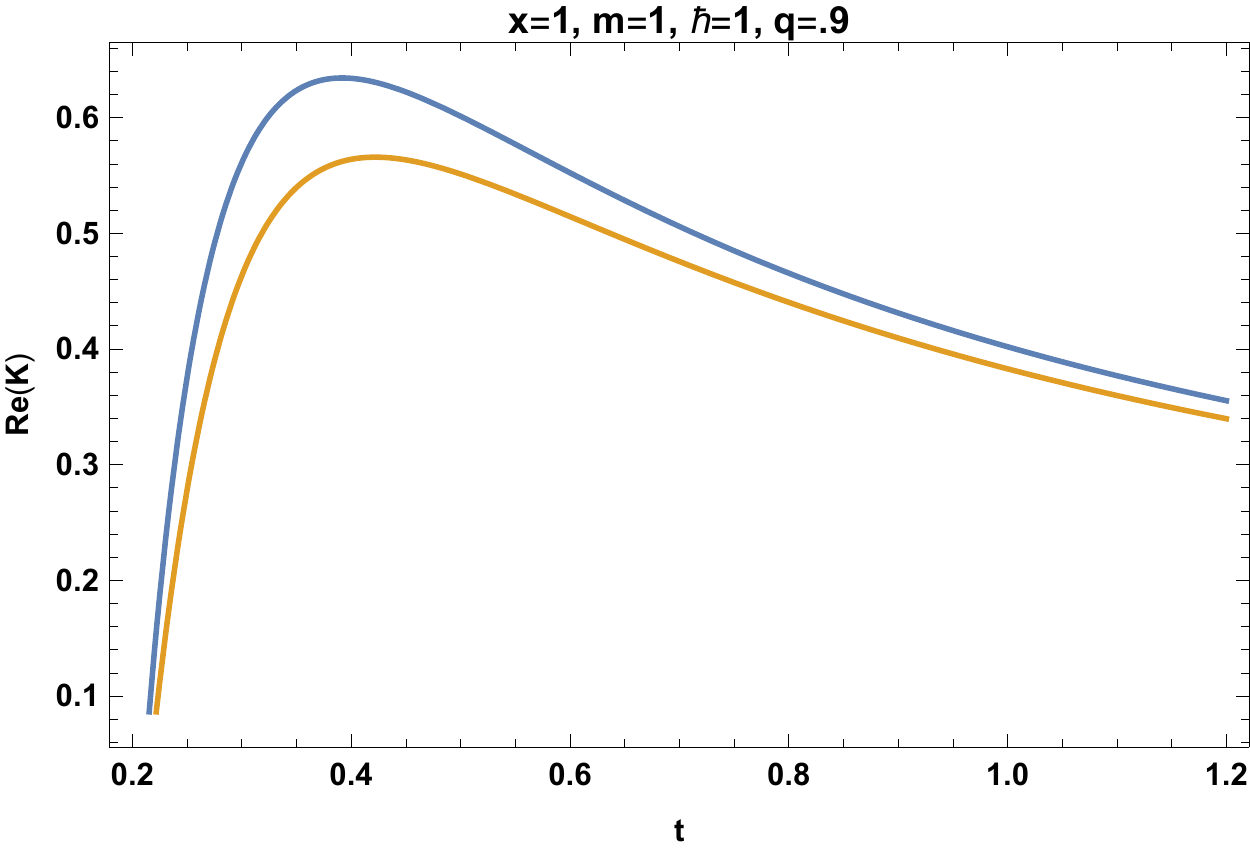}
\caption{Real parts of the modified propagator (blue line) vs. standard propagator (red line), for the free particle for the modified
statistics of Tsallis for $q=0.9$. We set the mass and the Planck constant to unity. Imposing $S_{cl} \lesssim \hbar$, it translates for fixed $x=1$ in $t \gtrsim1/2$,  and for fixed $t=1$ translates in $x^2 \lesssim 2$.}
\label{quantumT1}
\end{center}
\end{figure}

\section{The Harmonic Oscillator}
\label{oscProp}

In this Section we apply the formulation of a modified Quantropy of Section \ref{IQu} for the case of the harmonic oscillator.
We compute the modified propagator constructed by a superposition as previously. The extension
of quantum systems employing the modified q-statistics has only been made for the case of the free particle \cite{QMT}
with different arguments. 
Our proposal allows to search the manifestation of non-extensive statistics in quantum systems (non linear)
for generic potentials.  We illustrate the procedure calculating only $K_+$, the $K_-$
and $K_q$ cases could be similarly calculated.

\begin{figure}[h]
\begin{center}
\includegraphics[width=.23\textwidth]{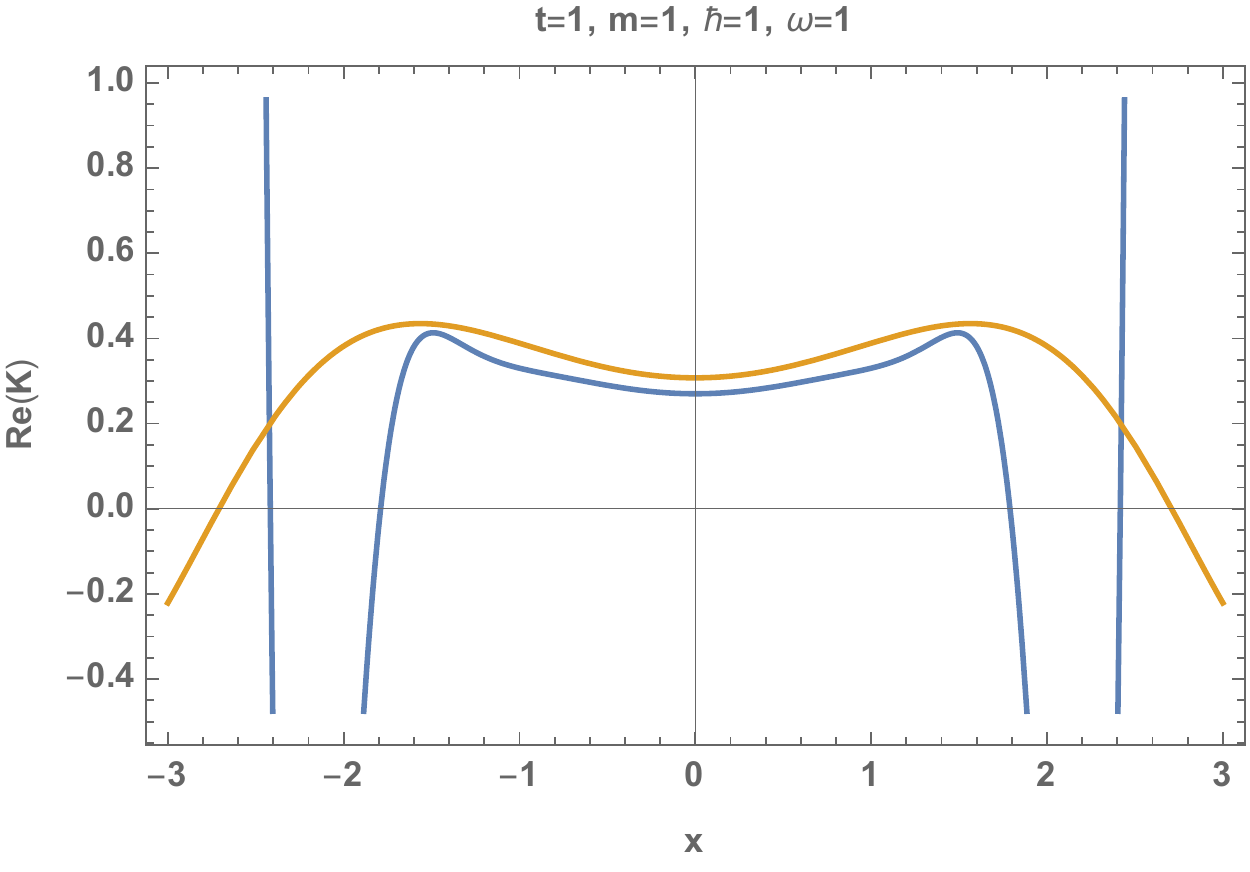}
\includegraphics[width=.23\textwidth]{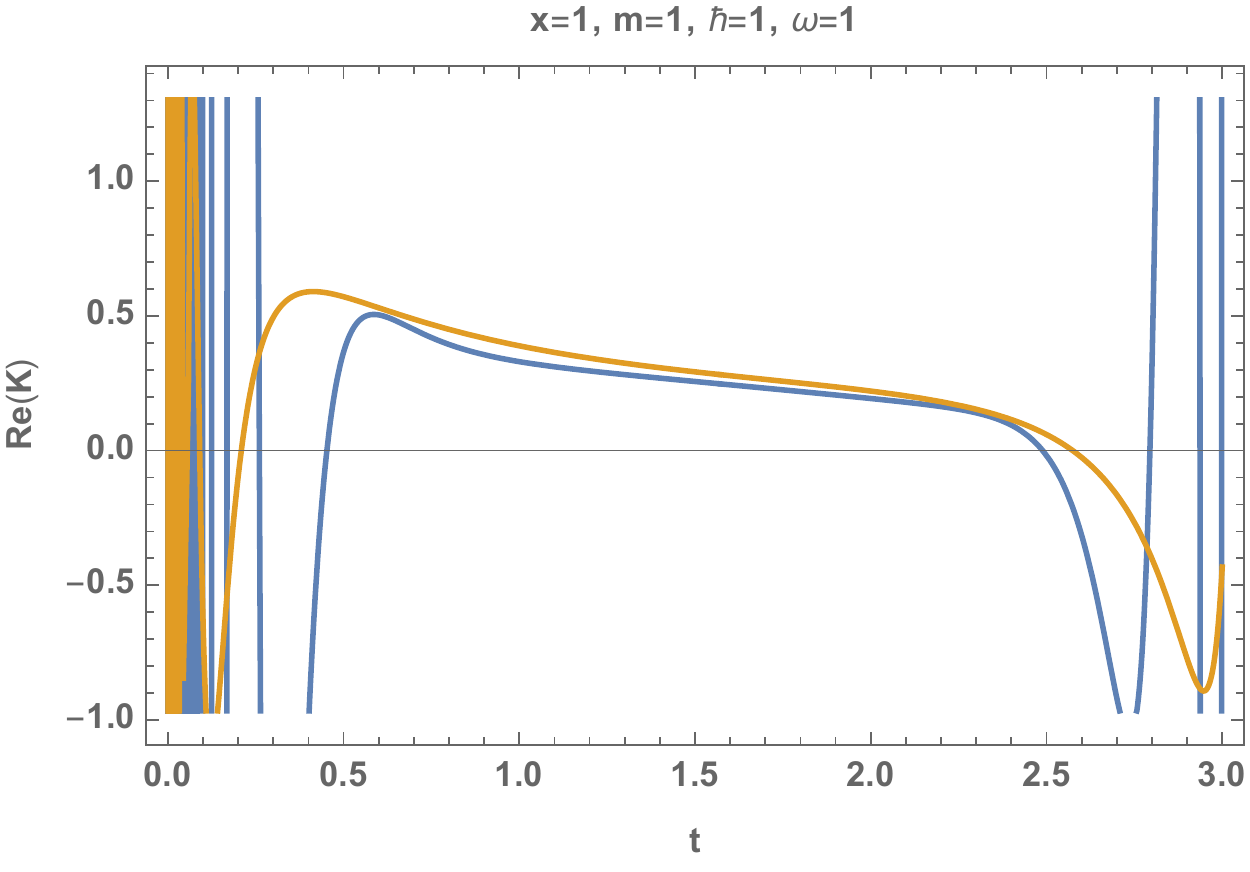}\\
\caption{We set the units $m=w=\hbar=1$. The left image shows the real parts of the normalized propagators for a fixed time $t=1$,
the region $S_{cl} \lesssim\hbar$ is given by $|x|\lesssim 1.765$. 
The {\it blue} line  corresponds to the modified propagator $K_+$,
meanwhile the {\it orange} one represents the usual propagator  $K(x,t,1)$. 
The right  plot is the comparison between the propagators amplitude for $x=1$
 where the {\it blue} and the {\it orange} is the correction and the usual respectively,
 the quantum region is given by $t\gtrsim 0.464$. }
\label{clas1}
\end{center}
\end{figure}

For the harmonic oscillator with Lagrangian $\mathcal{L}=\frac{m}{2}\dot x^2-\frac{m \omega^2}{2} x^2$ the path integral Kernel reads
\begin{align}
K(a,b)&=\left(\frac{m\omega}{2\pi i \hbar \sin \omega T}\right)^{1/2} \\
&\times \exp\left(\frac{i m\omega}{2\hbar  \sin \omega T} ( (x_a^2+x_b^2)\cos \omega T-2 x_a x_b)\right).\nn
\end{align}
Next, following a similar procedure as for the free particle we compute the generalized Kernel and normalize it. The unnormalized Kernel is given by:
\begin{equation}
K(x,t,1)= \exp\left(-\frac{1}{2} \lambda m \omega \cot (\omega t ) x^2 \right),
\end{equation} 
now we compute the next terms of the Kernel up to third order, those are:

\begin{align}
K(x,t,2)&= \left(\frac{\lambda}{2} \right)^2 \frac{\partial^2}{\partial \lambda^2} \exp(-2\lambda A) \nonumber \\  
&= \frac{1}{4} \lambda^2 m^2 \omega^2 x^4  \exp \left(-\lambda m \omega  \cot ( \omega t) x^2\right), \\
K(x,t,3)&=\frac{1}{2} \left(- \frac{\lambda^2}{3^2} \partial_{\lambda}^2 +2 \frac{\lambda^3}{3^3}\partial_{\lambda}^3 + 3 \frac{\lambda^4}{3^4} \partial_{\lambda}^4\right) \exp(-3 \lambda A) \nonumber \\ 
&=\left( -\frac{1}{8}  \lambda^2 m^2 \omega^2 \cot ^2(\omega t) x^4 \right. \nn \\
&\left. -\frac{1}{8} \lambda^3 m^3 \omega^3 \cot^3 (\omega t) x^6 \right. \nn\\
&\left.+\frac{3}{32}  \lambda^4 m^4 \omega^4 \cot^4 (\omega t) x^8 \right) \nonumber \\
&  \times \exp \left(\frac{-3}{2}  \lambda m \omega x^2 \cot (\omega t)\right) .
\end{align}
Thus the total normalized propagator up to third order reads

\begin{align}
&K_+(x,t) = \nn \\
&\frac{N_+}{\sqrt{2\pi}} \left\{ \left( \frac{ m \omega \cot(t \omega) }{i\hbar} \right)^{\frac{1}{2}} \exp \left(\frac{- m \omega x^2 \cot (t \omega)}{2  i \hbar} \right)  \right.  \\ 
&+ \frac{1}{4} \left( \frac{ m \omega \cot(t \omega) }{i\hbar} \right)^{\frac{5}{2}}  x^4 \exp \left(\frac{- m \omega x^2 \cot (t \omega)}{  i \hbar} \right) \nn \\
&- \frac{1}{8} \left[ \left( \frac{ m \omega \cot(t \omega) }{i\hbar} \right)^{\frac{5}{2}}  x^4  
+  \left( \frac{ m \omega \cot(t \omega) }{i\hbar} \right)^{\frac{7}{2}}  x^6 \right. \nn \\
&\left. \left. - \frac{3}{4} \left( \frac{ m \omega \cot(t \omega) }{i\hbar} \right)^{\frac{9}{2}}  x^8 \right] \exp \left(\frac{- 3m \omega x^2 \cot (t \omega)}{ 2 i \hbar} \right) +...\right\}, \nn
\end{align}
where $N_+ =\frac{1}{\sqrt{\cos(\omega T)}}\frac{1}{(1+\frac{3}{16\sqrt{2}}+\frac{1}{96\sqrt{3}})}$. The relative normalization of the modified to the usual propagator is a result obtained in Section \ref{freeProp},
see formulas (\ref{psiGen}-\ref{normA}). This result is universal i.e. independent of the action. In Figure \ref{clas1} we compare the propagator for the harmonic oscillator for the standard Quantropy and
for the one based on $S_+$ and $S_-$ statistics. We are interested in the quantum regime given by
$S_{cl}\approx \hbar$. There are noticeably effects in that regime. Outside the quantum region
oscillations grow as in the usual case \cite{FeynmanHibbs}. This behavior occurs in the classical region in which the modified 
Kernels will also not contribute.

\newpage

\section{Final Remarks}
\label{conclusion}

In this work we studied a newly proposed concept in the quantum theory inspired in
the entropy of statistical mechanics. It fills a previous gap on the analogy between both disciplines.
This is the main object developed by Baez and Pollard named {\it Quantropy}.
It can be regarded as an analytical continuation of the Entropy in Statistical Mechanics (S.M.)
to Quantum Mechanics (Q.M.); where probabilities of states change into complex amplitudes
of paths.  The identification performed is of energy in S.M. to action in Q.M., and from temperature to Planck constant,
it reads: $E\rightarrow S$ and $T\rightarrow i\hbar$.
We have considered an analogous definition at a macroscopic level with energy mapped to classical
action $E\rightarrow S_{cl}$, i.e.  considering  as the main entity instead of the amplitude $a(x)$ of the path
the propagator between space-time points $K(x)$. We constructed from the propagator
a kind of integrated version of the Quantropy $Q_0=-\int_X K(x)\ln K(x)$. In the standard BG case the maximization of the functional 
leads to $K_0\sim \exp(i S_{cl})$. This
functional can be generalized for modified entropies as $S_+$, $S_-$ and $S_q$ and the extrema of
it will lead to modified propagators. 

The result for the Tsallis statistics, leads to the propagator $K_q\sim \exp_q(i S_{cl})$,  this makes contact with the Tsallis result of modified wave function
for the free particle, because such propagator will lead to $\psi_q\sim \exp_q(i(k x-\omega t))$. The connection is made by arguments
of Feynmann and Hibbs \cite{FeynmanHibbs},  in the discussion of the propagator for the free particle $K_0\sim \exp(i S_{cl})$. 
They show that $K_0$ corresponds to the free particle wave function $\psi_0\sim \exp(i(k x-\omega t))$.  Thus analogously $K_q$
will lead to $\Psi_q$. As a further work we need to explore the relations in the case of the modified propagators $K_+$ and $K_-$. 
They will give rise to wave functions which
dependence are $\Psi_{\pm}=\exp_{\pm}(i(k x-\omega t))$, also for the free particle. In this case we have a recurrent series solution but we do not have exact expressions for these generalized exponentials.  As discussed
our proposal provides also generalized propagators  $K_+, K_-$ and
$K_q$ for problems with interactions; we illustrated this by considering the
$K_+$ associated with the harmonic oscillator.

There are hints  from previous studies that the modified entropies considered here can be interpreted as linked
with modified effective potentials. Therefore these modifications to the free particle could be
related to a usual quantum mechanics with an effective potential \cite{OGT}. However, this effects 
could also lead to non linear quantum equations explored in the literature with modified
wave functions \cite{TsallisQM1,TsallisQM2,QMT,Chavanis19}. Furthermore what we found here based on the concept of Quantropy,
could be linked to results for quantum systems in terms of usual entropy 
vs. the density-matrix \cite{CaboObregon18}. A system governed by a modified statistics ($S_+$, $S_-$ or $S_q$) will
lead to modified density-matrix distributions. 

Moreover the modified ``propagators'' $K_+,K_-$ and $K_q$ actually are strictly no longer standard propagators,
because they lack the usual propagation property. This means that is not equivalent to propagate the particle from 
$(0,0)$ to $(t_2,x_2)$, than to first propagate it from $(0,0)$ to $(t_1,x_1)$ and then from $(t_1,x_1)$ to $(t_2,x_2)$.
This is seen because for example $\int K_+(0,0;x_1,t_1)K_+(x_1,t_1;x_2,t_2) dx_1 \neq K_+(0,0;x_2,t_2)$. 
This relates to the fact that in a quantum open systems where this generalized entropies are motivated, 
the nature of the processes are Non-Markovian. 
Those systems in consideration are modeled with Master Equations(Stochastic) \cite{RevModPhys.89.015001}. 

We would like to further explore processes where the modified statistics in Quantropy play a central role. This could be done
via the implied modified wave functions, which could also be interpreted as coming from standard quantum mechanics
with an effective interaction or from non-linear quantum equations. The modified wave functions will correspond to the modified propagators  here obtained.  One should then explore in detail their quantum mechanic evolution. For some physical systems, in particular the harmonic oscillator, the $K_+$, $K_-$ and $K_q$ will illustrate its modified quantum behavior.
That is the matter of future work.

\section{Acknowledgments}

We thank Alejandro Cabo, Vishnu Jejjala, Oscar Loaiza-Brito, Miguel Sabido, Marco Ortega, Nels\'on Flores-Gallegos and Pablo L\'opez-V\'azquez  
for useful discussions and comments. NCB thanks PRODEP NPTC UGTO-515  Project, CIIC 181/2019 UGTO Project
and CONACYT Project A1-S-37752. RSS thanks to CONACYT and PRODEP NPTC UDG-PTC-1368 Project for supporting this work. 
OO thanks CONACYT Project 257919 and CIIC 188/2019 UGTO Project.

\appendix

\section{Numerical Approach}
\label{app}
\begin{figure}[h]
\begin{center}

\includegraphics[width=.4\textwidth]{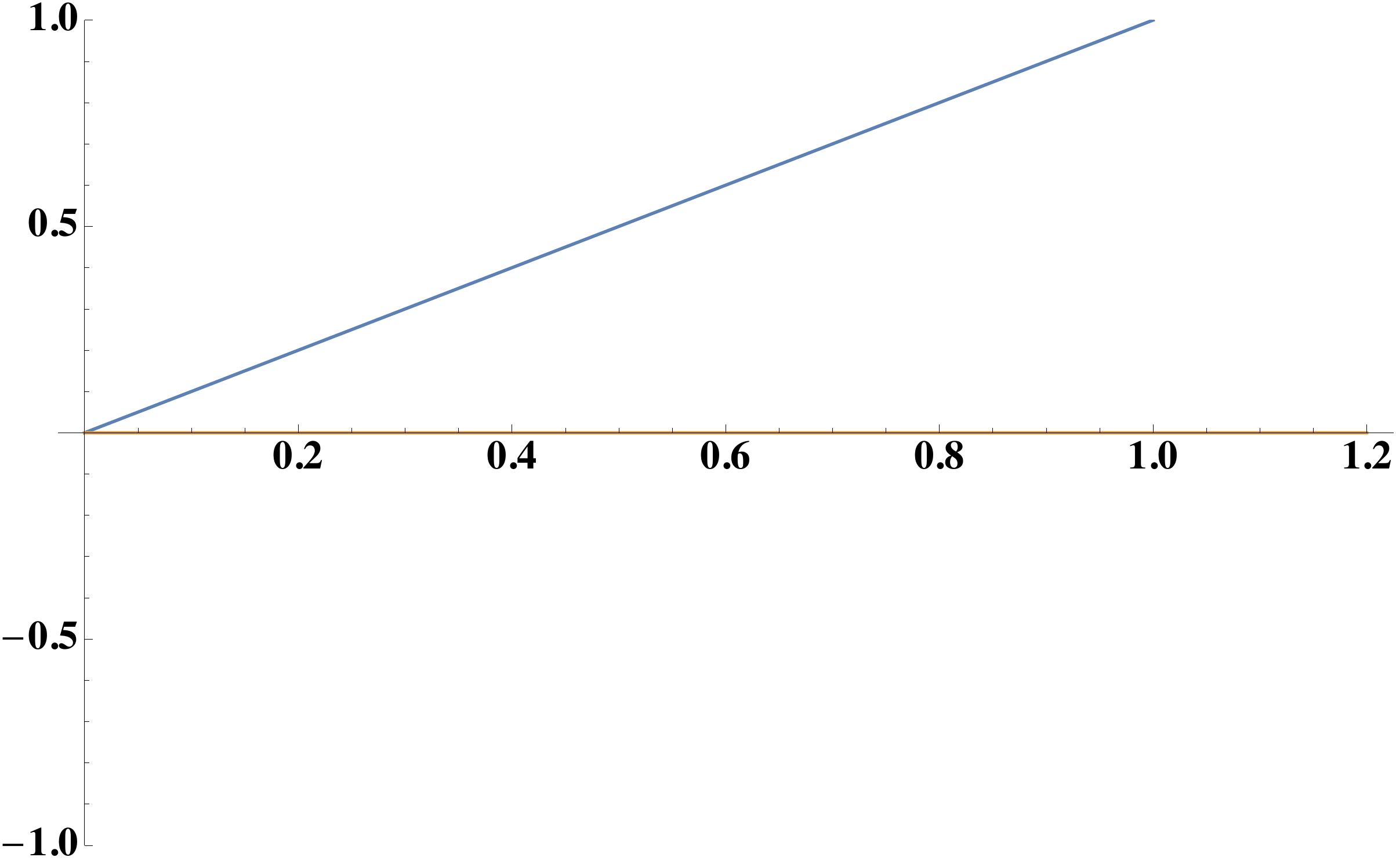}\\
\caption{The plot represents the check of the numerical solution for $K_+$. The yellow line $\text{Re}(F (K(\frac{S_{cl}}{i\hbar})))$
should be zero and the blue line $\text{Im}(F (K(\frac{S_{cl}}{i\hbar})))$ the identity function, in order to satisfy 
equation (\ref{ecuI}) with $A\rightarrow S_{cl}$ and $a\rightarrow K$.
 }
\label{fig:check}
\end{center}
\end{figure}

In this appendix we shall develop a numerical approach to study the propagator $K_+$ vs. the action $S_{cl}/\hbar$.  To construct a relation between the action and the Kernel it is proposed an interpolating function which is forced to satisfy (\ref{extrema}). This is, given a real value of the action ($S_{cl}/\hbar$), the interpolating function, finds a complex value of the Kernel that satisfies  (\ref{extrema}). The numerical function is expected to improve the convergence for larger values of the action with respect to the exponential expansion proposed in (\ref{amplitud}). In Figure \ref{fig:check} we show the plot obtained thorough the numerical approach. The X axis represents the numerical value of the action, and the Y axis represents the real (yellow line) and imaginary (blue line) parts of the function $F (K(\frac{S_{cl}}{i\hbar}))=-i S_{cl}/\hbar$ in (\ref{ecuI}).

\begin{figure}[!h]
\begin{center}
a) \includegraphics[width=.4\textwidth]{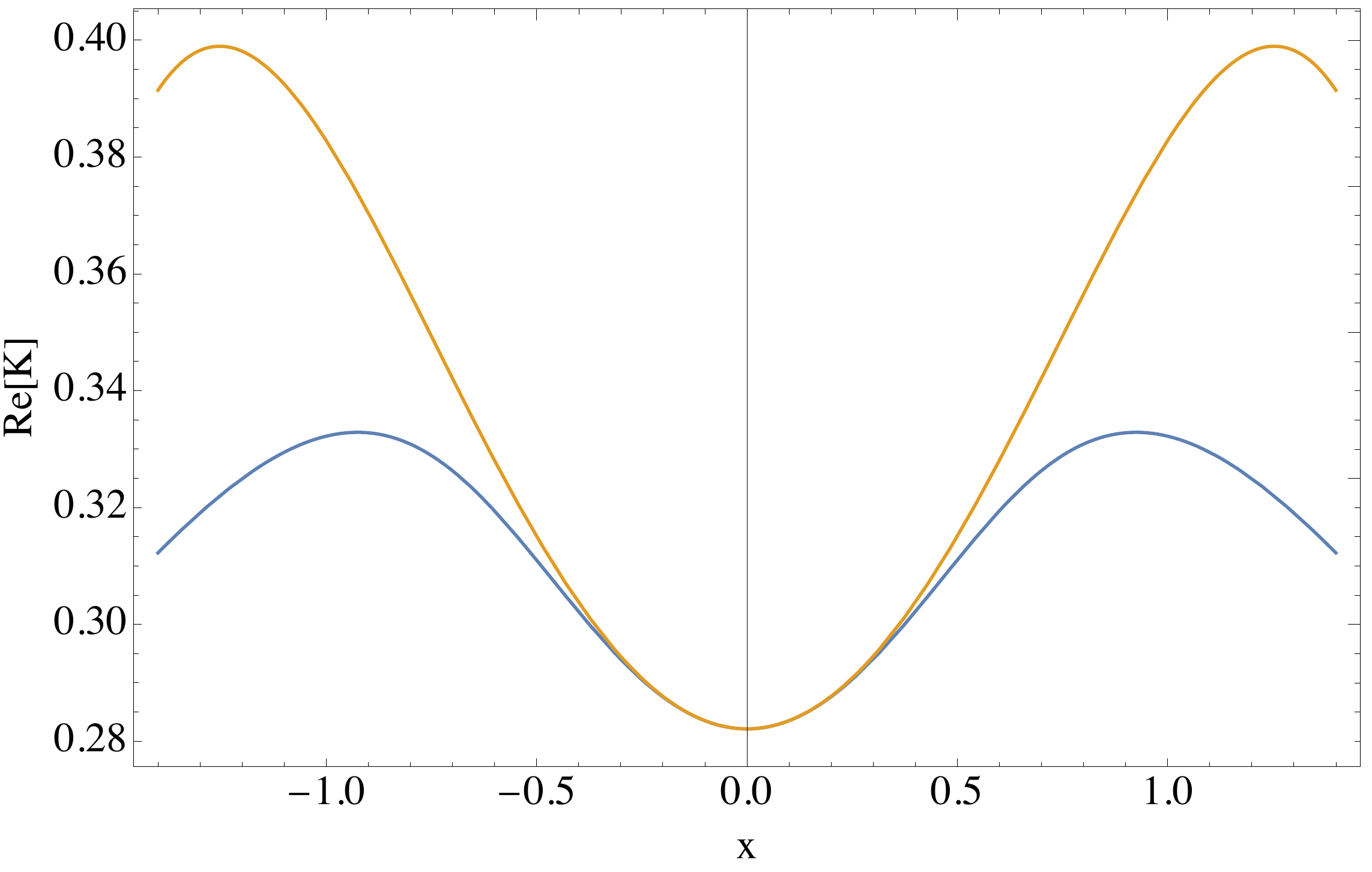} \\ b) \includegraphics[width=.4\textwidth]{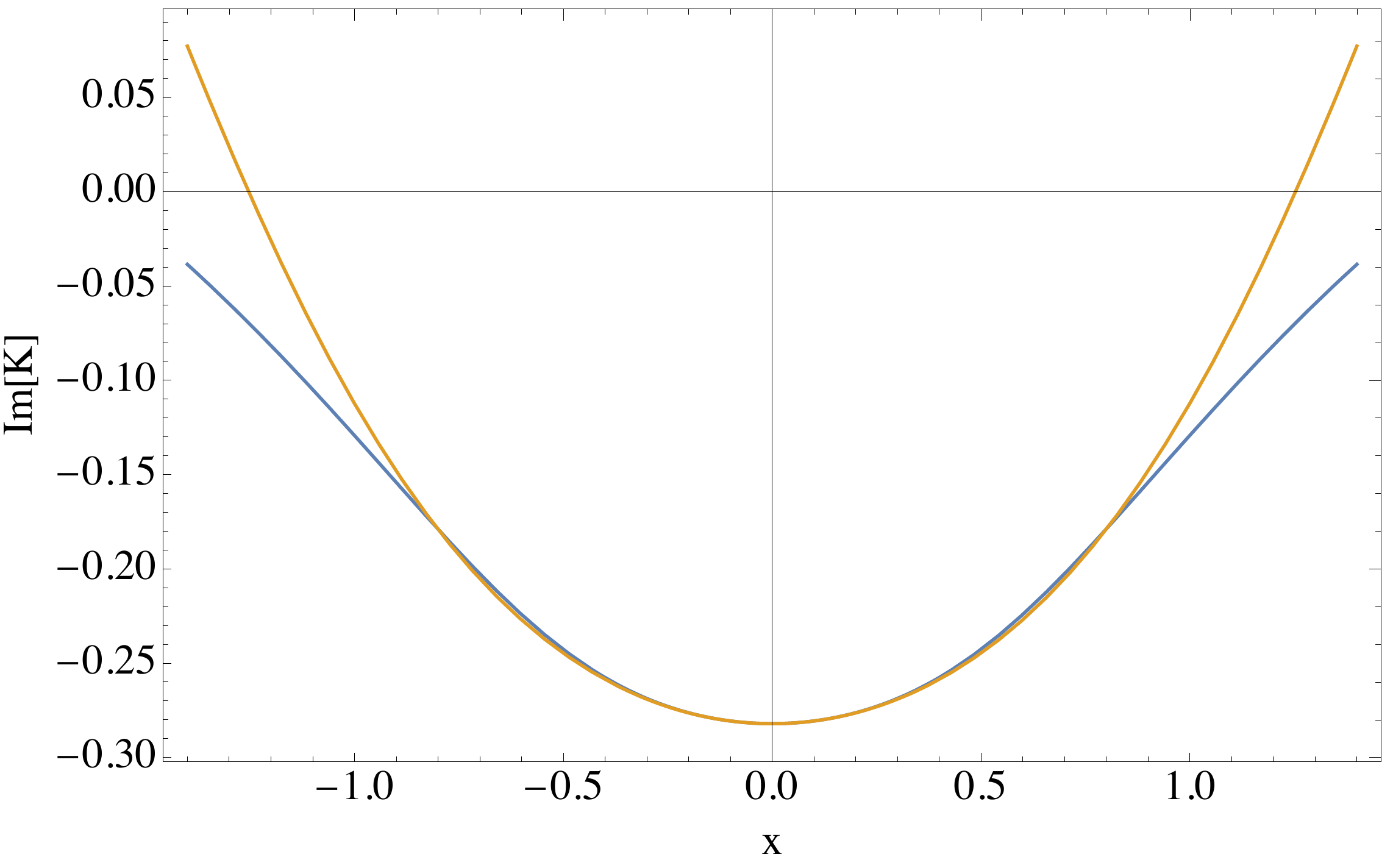}\\
\caption{Numerical check for the series expansion of the modified free particle Kernel. We consider the units $m=t=\hbar=1$. The blue line corresponds to the modified Kernel whereas the yellow line is related to the standard Kernel. }
\label{fig:numapp}
\end{center}
\end{figure}

For the case of the free particle the results are shown in Figure \ref{fig:numapp},  
we show the imaginary and real part of the Kernel. 
The numerical function denoted is represented as $exp_+ $ which is the behavior found from the quantropy introduced in Section \ref{IQu}. Thus, the free particle Kernel takes the form
\begin{equation}
K_+ \sim \left( \frac{m}{2 \pi i \hbar t} \right)^{1/2} \exp_+ \left( i \frac{m x^2}{2 \hbar t^2} \right). \,
\end{equation}
Using this definition we are able to compute the Kernel in a numerical manner. In Figure \ref{fig:numapp} we present the modified versus the usual Kernel. Thus as is observed the modified Kernel has a similar behavior that the standard Kernel for $S_{cl}<<\hbar$, but it deviates in  already
in the quantum regime. However, since the free particle wave function is not normalizable we shall not proceed further to check the probability distribution.\\

\begin{figure}[!h]
\begin{center}
a) \includegraphics[width=.4\textwidth]{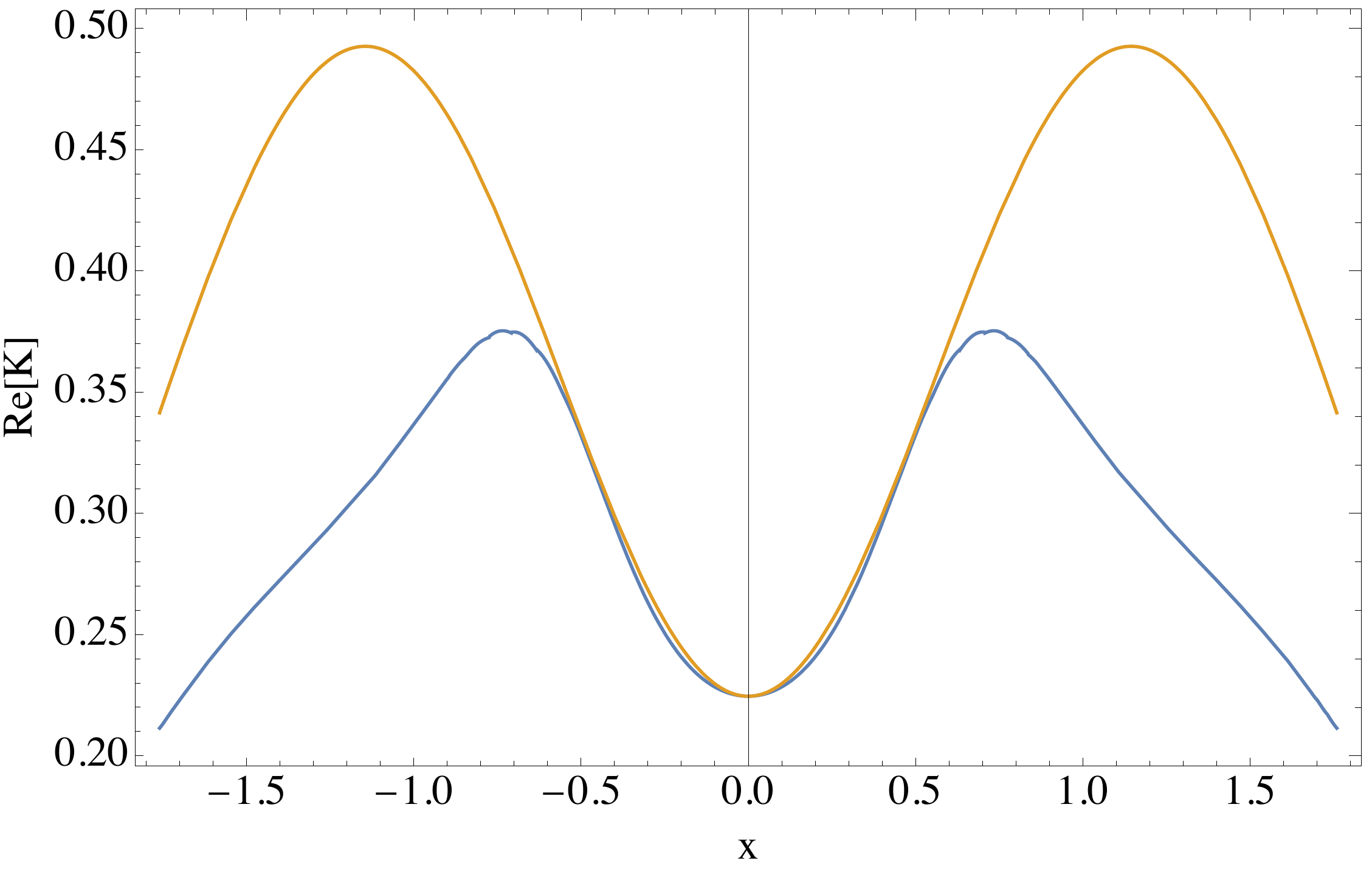} \\ b) \includegraphics[width=.4\textwidth]{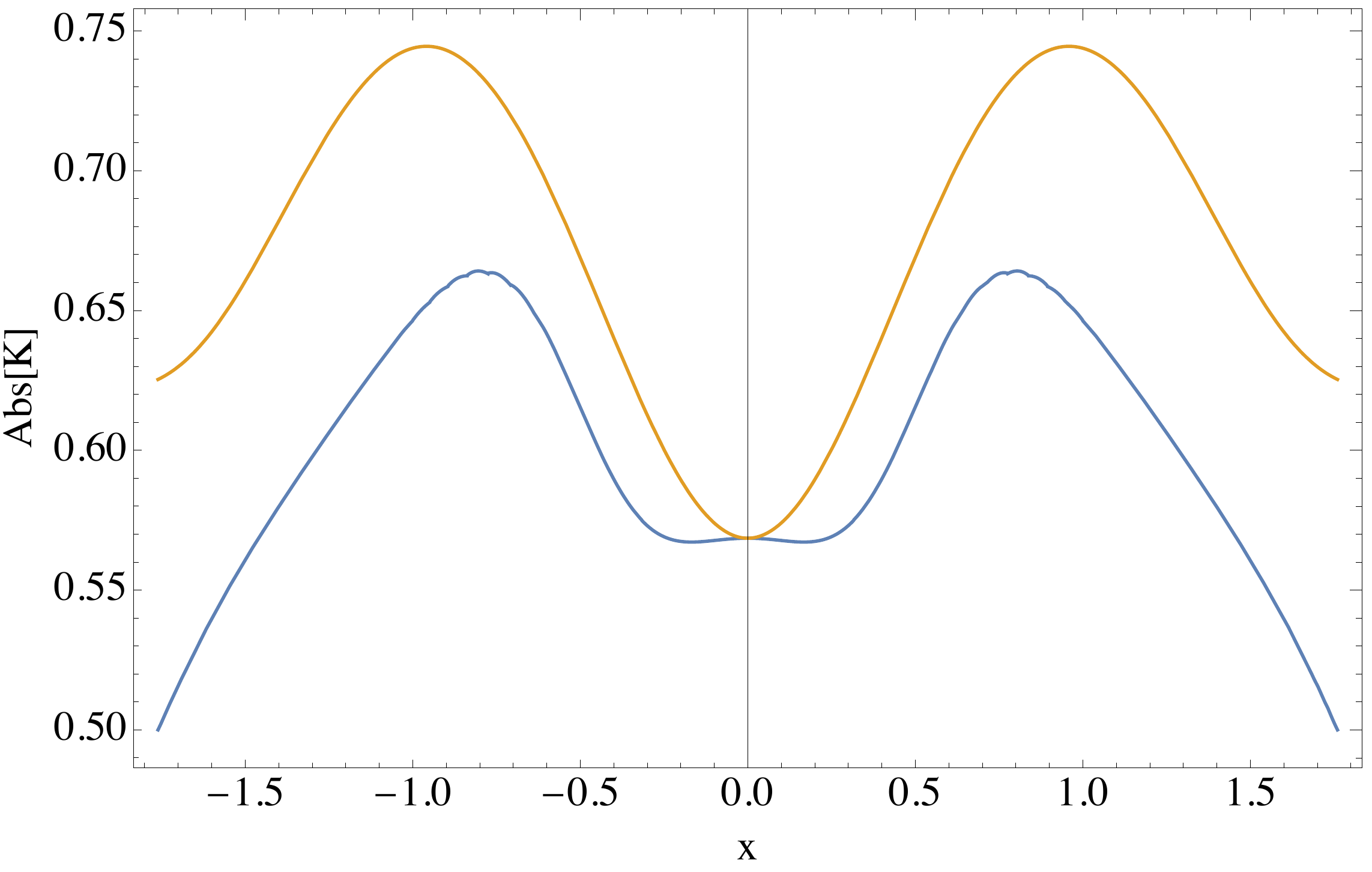}\\
\caption{Numerical check for the harmonic oscillator propagator. We consider the units $m=\omega=t=\hbar=1$. The blue line corresponds to the modified Kernel whereas the yellow line constitutes the standard Kernel. }
\label{fig:numapp2}
\end{center}
\end{figure}

Now, let us proceed to check the numerical approach of the Kernel for the case of the harmonic oscillator. In order to match with the standard results known for the harmonic oscillator we shall employ the expansion
\begin{equation}
K =\sum_{n=0}^\infty \exp \left( -\frac{i}{\hbar} E_n T \right) \phi_n (x_b) \phi_n^\ast (x_a) \,,
\end{equation}
where as usual
\begin{align}
\phi_n (x) &= \frac{1}{(2^n n!)^{1/2}} \left( \frac{m \omega}{\pi \hbar} \right)^{1/4} \nn\\
&\times H_n \left( x \sqrt{\frac{m \omega}{\hbar}} \right) \exp \left( -\frac{m \omega}{2 \hbar}x^2\right) \,,   
\end{align}
and $H_n$ are the Hermite polynomials defined by the generating function
\begin{equation}
H_n (y) = (-1)^n \exp\left( y^2 \right) \frac{d^n}{d y^n}\exp \left( -y^2 \right) \,.
\end{equation}
For the case of the numerical function, the Kernel is written in a perturbative form as
\begin{align}
K_+& = \left( \frac{m \omega}{\pi \hbar} \right)^{1/2} z^{-1} p_1\nn \\
&\times \exp_+ \left( -\frac{m \omega}{2 \hbar} p_2 \left[ x_a^2 + x_b^2 \right]+ 4 p_3 x_a x_b \right), 
\end{align}
where $p_i$ are the Taylor polynomials of the functions
\begin{align}
p_1& = \left( \frac{1}{1-z^2} \right)^{1/2} \,, \quad p_2 = \left( \frac{1+z^2}{1-z^2} \right)^{1/2} \,, \nn \\
& p_3 = \left( \frac{z}{1-z^2} \right)^{1/2} \,.
\end{align}
As in the free particle case the modified entropy shall provide a new function $\exp_+$. In Figure \ref{fig:numapp2} we show the comparison of the usual propagator and the modified propagator employing the numerical function.

For the harmonic oscillator, the modified Kernel deviates from the standard Kernel in the quantum region.

\bibliographystyle{utphys}
\bibliography{biblio4.bib}

\end{document}